\documentclass[useAMS,referee,usenatbib]{mn2e}  
\usepackage{epsfig}  
\usepackage{graphicx,rotating,amssymb,supertabular,longtable,lscape}
\usepackage{amsmath}
\usepackage{textcomp}

\def\apj{\mbox{ApJ}}
\def\apjl{\mbox{ApJL}}
\def\apjs{\mbox{ApJS}}

\def\mnras{\mbox{MNRAS}}
\def\aj{\mbox{AJ}}

\def\nat{\mbox{Nature}}
\def\aap{\mbox{A\&A}}

\newcommand\mums{\textmu m }
\newcommand\mum{\textmu m}
   
\title[A Dust Emission Model of AGN]
{Revisiting the Infrared Spectra of Active Galactic Nuclei with a New Torus 
Emission Model}
       
\author[Fritz et al.]{J. Fritz$^{1,2}$\thanks{Models publicly available, contact:
fritz@pd.astro.it}, A. Franceschini$^{1}$, E. Hatziminaoglou$^{2}$\\
$^{1}$Dipartimento di Astronomia, Vicolo Osservatorio 2, I-35122 Padova, Italy\\
$^{2}$Instituto de Astrof\'{i}sica de Canarias, C/ V\'{i}a L\'{a}ctea s/n, E-38200 La Laguna, Spain}
  
\begin{document}       
  
\maketitle  
  
\begin{abstract}   
\noindent
We describe improved modelling of the emission by dust in a toroidal--like structure
heated by a central illuminating source within Active Galactic Nuclei (AGN).
We chose a simple but realistic torus geometry, a flared disc, and a
dust grain distribution function including a full range of grain sizes. 
The optical depth within the torus is computed in detail taking into account 
the different sublimation temperatures of the silicate and graphite grains,
which solves previously reported inconsistencies in the silicate emission feature 
in type-1 AGN.
We exploit this model to study the spectral energy distributions (SEDs) of $58$ 
extragalactic (both type-1 and type-2) sources using 
archival optical and infrared (IR) data. We find that both AGN and starburst 
contributions are often required to reproduce the observed SEDs, although in 
a few cases they are very well fitted by a pure AGN component. The AGN contribution to 
the far-IR luminosity is found to be higher in type-1 sources, with all the type-2 
requiring a substantial contribution from a circum-nuclear starburst.
Our results appear in agreement with the AGN Unified Scheme, since the distributions 
of key parameters of the torus models turn out to be compatible for type-1 and type-2 AGN. 
Further support to the unification concept comes from comparison with medium-resolution IR spectra of type-1 AGN by the Spitzer observatory, 
showing evidence for a moderate silicate emission around $10$ \mums which our code reproduces.
From our analysis we infer accretion flows in the inner nucleus of local AGN
characterized by high equatorial optical depths ($A_V\simeq 100$), 
moderate sizes ($R_{max}<100\ pc$) and very high covering factors 
($f\simeq 80$ per cent) on average.

\end{abstract}  
  
\maketitle  
     
\section{INTRODUCTION}
\label{intro}         

The idea that the quasar infrared (IR) spectral energy distribution (SED) could be 
dominated by thermal emission from dust was already proposed by \cite{neugebauer}, 
who noted an excess at $\sim 3.5$ \mums with respect to the power--law emission that was 
known to dominate the ultraviolet (UV) and optical spectrum. 
\cite{barvainis} was the first to demonstrate that it was possible 
to describe this emission by means of thermal radiation from dust heated by 
the primary optical/UV continuum known to dominate type-1 AGN SED. He considered 
an optically thick, either smooth or a clumpy dust distribution within a 
non-spherical geometry. Such a model could naturally account for the observed $3$ 
\mums bump, which was attributed to graphite grains with temperatures close to 
their sublimation limit. The broadness of the IR SED, that could not be reproduced 
by means of a single-temperature black body, was explained in terms of multiple 
temperature components of dust. 

More quantitative models have been developed since then, accounting for different 
features and using different computational approaches for solving the radiative transfer 
equation. \cite{pierkrolik} suggested an emission model from an 
annular ring of dust. Assuming a common extinction curve for all of the grains, 
a distribution function of grain composition and sizes as given in \cite{MRN} 
and not accounting for light scattering, they solved the radiative transfer 
problem with the B\"ohm-Vitense method described in \cite{mihalas78}. Based on this model, they explored 
the main properties of IR emission in AGN, such as the width of the SED and 
the wavelength of peak emission, IR colours and the presence and/or absence 
of the $9.7$ \mums silicate feature. 

\cite{granato94} solved the radiative transfer equation by means of the 
{\it $\Lambda$--iteration} method (see \ref{sec:lambda-iter} for more details), 
adopting the \cite{rr92} model for interstellar dust, based on both graphite and 
silicate grains, and taking into account radiation scattering. Their adopted 
geometry was the so--called {\it flared disc}, with dust grain density that was 
allowed to vary both along the radial and the vertical coordinates. 

\cite{stenholm} studied a configuration given by an optically thick silicate disc, 
finding no need for silicate grains depletion or non-standard dust composition in 
order to weaken the strength of the $9.7$ \mums feature, which turned out to 
be weak in most of his models with optical depth larger than $1$. Furthermore he found 
that, in contrast with \cite{pierkrolik}, scattering is not negligible and does 
affect the shape of the near IR emission in those models where optical depth is high.

\cite{efstat95} analysed three different geometrical configurations for the torus,
a flared and a tapered disc and an anisotropic sphere, and found that a tapered disc (i.e. 
a disc whose height increases along with distance from the center but flattens at a 
constant height in the outer parts) with an opening angle of $45^\circ$ was the most 
succesful representation of what was observed. They followed 
a ray--tracing numerical method, with the same dust mixture as in \cite{rr92}. 
The sublimation temperature was assumed to be $1000$ K for all grains, while the 
distances from the central source where this temperature is reached were
assumed to be dependent on the grain size.
In the same line,
\cite{Manske98} modelled the dusty torus as a flared disc in which both graphite 
and silicate where taken to sublimate at a temperature of $1500$ K. In their analysis 
they concluded that flared tori, dust composed of silicate and graphite 
grains and a power-law density distribution failed to explain the observed 
featureless AGN spectra.

In parallel, \cite{nenkova02} developed a torus model in which
dust was distributed in clumps. They solved the radiative transfer for each
clump by means of the radiative code {\sc dusty} \citep{nenkova99} and then
computed the final emission as the sum of clumps' emission calculated using
a ray--tracing approach.

At variance with \cite{efstat95}, \cite{bemmel-dullemond} claim that the 
obscuring torus could not be flat (tapered, for example), and should rather be conical or 
flaring. Their more reliable models predicted constant 
density over the whole torus and a quite large value for the inner radius 
($\sim 10$ pc), which they found necessary in order to fit observed colours. 
They also tried to modify the power-law distribution function 
for grain sizes with respect to that given by \cite{MRN}, in order to 
attenuate the strength of the $10$ \mums silicate feature and recently
concluded that there was no need for a clumpy dust distribution \citep{dullemond-bemmel}.


In the present paper we try to combine most of the successful features of the previous works to
built an improved model explaining IR emission in AGN. The paper is organised as follows.
Section \ref{model} presents the physical and computational details of the model.
Section \ref{param} discusses the contribution of each of the model parameters in the
produced SEDs. In Section \ref{fit} we explain the observed SEDs of a series of objects 
taken from the literature using various realisations of the torus model in combination
with starburst templates. Finally in Section \ref{discuss} we discuss our results and conclusions.

\section{Model Description }
\label{model}

\subsection{Torus Geometry and Grain Sublimation}\label{model1}

We use a simple but realistic geometry for the spatial
distribution of dust: the {\it flared disc} (\citealt{efstat95}, 
\citealt{Manske98} and \citealt{bemmel-dullemond}), that can be represented 
as two concentric spheres, delimiting respectively the inner and the outer 
torus radius, having the polar cones removed.

\begin{figure}
\centering
\includegraphics[height=0.36\textwidth]{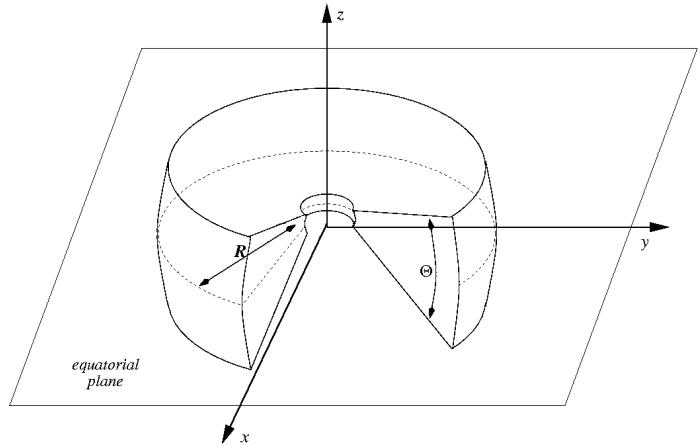}
\caption{Shape of the toroidal dust distribution in our model. In this figure $R$ is 
the ratio of the external to internal radius and $\Theta$ is the full--opening 
angle of the torus.}
\label{fig:torus}
\end{figure}

The size of the torus is defined by the outer radius $R_{max}$ -- the inner radius being 
defined by the sublimation temperature of dust grains under the influence
of the strong nuclear radiation field  -- and by the angular 
opening angle $\Theta$ of the torus itself (see Fig. \ref{fig:torus}). 
%
%
Considering a typical graphite grain with radius 
$a_G=0.05$ \mums and using the thermal equilibrium equation 
(see Eq. \ref{eqn:thrm}), \cite{barvainis} derives the following 
formula to compute the distance at which this grain reaches the 
sublimation temperature:
\begin{equation}\label{eqn:rmin}
R_{min}\simeq 1.3 \cdot \sqrt{L_{46}^{AGN}}\cdot T_{1500}^{-2.8} \qquad [pc] ,
\end{equation}
where $L_{46}^{AGN}$ is the bolometric ultraviolet/optical luminosity 
emitted by the central source, expressed in units of $10^{46}$ erg s$^{-1}$ and 
$T_{1500}$ is the sublimation temperature of the dust grain given in 
units of $1500$ K.
Note that this relations depends on both the grain dimension and species
but we here assume as a common minimum radius the average one. We account for
the different sublimation temperature of silicate grains, however, yielding
a minimum radius that can be up to three times larger than the one for graphite.
Furthermore, the presence of the dust layer composed by graphite grains,  
that attenuates the X-ray and UV emission, is taken into account when computing the
silicate minimum radius.

\subsection{Definition of the geometrical grid}
\label{sec:geomgrid}

\begin{figure}
\centering
\includegraphics[height=0.41\textwidth]{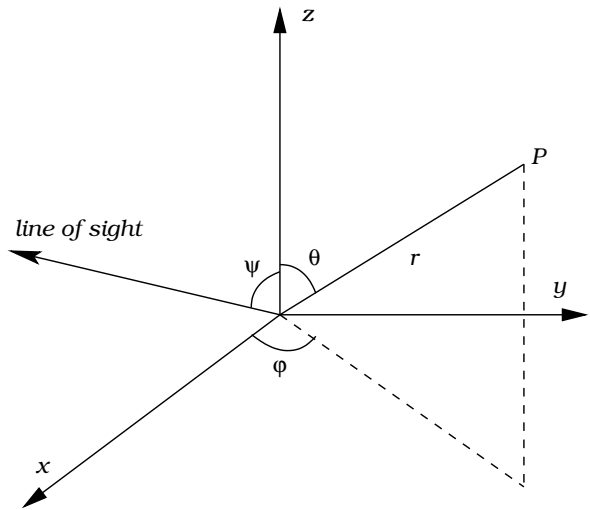}
\caption{The geometrical grid is defined with respect to a spherical coordinates system.
The equatorial plane of the torus lies in the {\it xy} plane, so that when 
the view angle $\Psi=90^\circ$ the geometry reproduces a type-2 objects, while 
$\Psi=0^\circ$ would correspond to the type-1 view, according to the 
Unified Model.}
\label{fig:sys_ref}
\end{figure}

%
%
We define a geometrical grid with reference to a polar coordinates system
$(r,\theta,\varphi)$, as shown in Fig. \ref{fig:sys_ref}, 
by dividing the torus in {\it volume elements}. We then compute all the 
quantities of interest within a given volume element as if all dust properties
were those of the center.
Since the temperature gradient is higher in the inner parts, we used a
logarithmic--spaced division for the radial coordinate, in 
order to better sample the properties of the emission in the 
innermost regions. 

The grid has a
constant number of subdivisions in the azimuthal coordinate ($\varphi$)
that is fixed to $36$, while the equatorial distance coordinate is divided in $12$,
$14$ and $16$ parts for models with respectively $60^\circ$, $100^\circ$ and
$140^\circ$ of torus amplitude. The radial coordinate is divided
in $40$, $80$ and $120$ parts for $R_{max}/R_{min}=30$,
$100$ and $300$ respectively. The wavelentgh values at which the
spectrum is computed consist in a grid equally-spaced in logarithm
with $120$ points in the range from $10^{-3}$ to $10^3$ \mum.

Neither the computation of the emitted radiation from each grain species in 
each volume elements within the torus, nor the computation of the optical 
depth between all torus cells are particularly CPU time consuming. However,
for a typical $\sim 15000$ element grid (see Sect. 3.1 below)
this computation must be repeated $\sim 15000^2$ times, if we want to 
evaluate the contribution to the radiation field in each element from
all other in the torus.
It is therefore essential to exploit all the symmetries of the adopted 
geometry, identifying groups of volume elements whose properties are identical all
over the torus grid. For instance, all volume elements at any given
$\varphi$ have exactly the same dust density, temperature and emission 
conditions.
It is sufficient to compute these quantities just for these elements, which we 
will refer to as {\bf sample elements}. 

Furthermore, when computing the incoming radiation on a given sample element, 
the problem is symmetric with respect to the $xz$ plane, so it is possible 
to compute incoming radiation from just one side of the torus.

\subsection{Dust properties}

The main dust components are silicate and graphite grains, 
in almost equal percentages.  The former are responsible for the
observed absorption feature at $\sim 9.7$ \mums in type-2 objects, 
while the latter are responsible for the rapid 
decline of the emission at wavelength shortwards of a few \mum, 
corresponding to a black body emission of about $1500$ K, the sublimation
temperature for these particular grains. 

Most models in the literature make use of a small number of grain categories
at different sizes (e.g. \citealt{granato94}; \citealt{efstat95}). 
Following an approach similar to that of \cite{bemmel-dullemond}, 
we decided to adopt a more complete distribution using, 
for each of the two species, scattering and absorption coefficients given
by \cite{laor_draine93} for different grains sizes  from $0.005$ to $0.25$ \mums
and from $0.025$ to $0.25$ \mums for graphite and silicate respectively.

As for the distribution function for the grains sizes $a$ we adopted the MRN 
distribution (see \citealt{MRN}):
\begin{equation}\label{eqn:grains}
dN(a)=10^{A_i}\cdot a^{q} da
\end{equation}
where $dN(a)$ is the number of grains with radius between $a$ and $a+da$ 
normalized to the number of hydrogen atoms. As for the exponent $q$, 
the value of the standard galactic extinction curve 
(\citealt{MRN}, $q=-3.5$) has been adopted, so that the extinction curve
that characterizes our model is similar to the galactic one.
The constant $A_i$ gives the normalization with respect to hydrogen 
abundance, and it is taken to be $-25.16$ for graphite and $-25.11$ 
for silicates \citep{drainelee84}.

\subsection{The density law throughout the torus}

We adopted a law for the gas density within the torus allowing for a density 
gradient along the radial and the polar distance ($\theta$) coordinates:
\begin{equation}\label{eqn:rho}
\rho\left(r,\theta \right)=\alpha\cdot r^{\beta}\cdot e^{-\gamma\times |cos(\theta)|} .
\end{equation}
The normalization constant $\alpha$ here is a linear function of the equatorial 
optical depth (see Section \ref{sec:compute}). The latter will be used in the following instead of the 
$\alpha$ parameter.

\subsection{Spectral intensity of the central power source}

We assume that the torus is illuminated by a central point-like energy 
source with isotropic emission. We define its spectrum in the wavelength range 
$0.001$ \mums $\div 20$ \mums (corresponding to the frequency range 
$\sim 3\cdot 10^{17} \div \sim 1.5\cdot 10^{13}$ Hz) 
and described it as a composition of power laws with variable indices. 
The values that we adopted are the same as in \cite{granato94} or 
\cite{nenkova02}, namely:
\begin{equation}\label{eqn:agn}
\lambda \cdot L(\lambda)=\left\{
\begin{array}{lrrr}
L_0\cdot \lambda^{1.2}  & [erg/s] & \mbox{if} \; 0.001 < \lambda < 0.03  & [\mu]\\
L_0\cdot \lambda^{0}    & [erg/s] & \mbox{if} \; 0.03  < \lambda < 0.125 & [\mu]\\
L_0\cdot \lambda^{-0.5} & [erg/s] & \mbox{if} \; 0.125 < \lambda < 20.0  & [\mu]
\end{array}
\right.
\end{equation}
where $L_0$ sets the bolometric luminosity normalization constant.

\subsection{Solution of the radiative transfer equation}\label{sec:lambda-iter}

The radiation emitted by the above described physical configuration is
given by the sum of the primary source located in the torus center 
and a secondary contribution given by 
thermal and scattering dust emission. 
Therefore dust plays a double role, first absorbing 
the radiant energy emitted by the primary source, partially or totally 
obscuring it, then re-emitting it at longer wavelengths, 
typically in the spectral range $1\div 1000$ \mum. 
A complication arises from the fact that dust can be optically 
thick to its own radiation. 
This requires the solution of the radiative transfer equation 
[Eq. (\ref{eqn:transfer}) below]. 
In particular the presence of a dust scattering component makes this 
equation not solvable analytically: the source function 
is not known {\it a priori}, depending on the solution $I_\nu$ itself. 

The thermal and scattered components of dust radiation 
are described by the absorption ($\alpha_\nu$) 
and the scattering ($\sigma_\nu$) coefficients respectively, 
so that the radiative transfer equation will be written as:
\begin{equation}\label{eqn:transfer}
\frac{dI_\nu}{ds}=-\left(\alpha_\nu+\sigma_\nu\right)\cdot \left(I_\nu -S_\nu \right)
\end{equation}
in which $S_\nu$ is the {\it source function}, that can also be written as
(see \citealt{rybickilight}):
\begin{equation}\label{eqn:source}
S_\nu=\frac{\alpha_\nu B_\nu + \sigma_\nu J_\nu}{\alpha_\nu+\sigma_\nu} , 
\end{equation}
i.e. a weighted average of the two separate source functions. 

We solved the transfer equation numerically adopting the 
{\it $\Lambda$--iteration} method: 
A lower limit for the incident radiation is estimated assuming that the
only source of radiation (and hence of dust--heating) is the central, 
non-thermal, source. 
The equilibrium temperature (and hence 
thermal emission) of the grains of each species in each sample 
element is found by solving the thermal equilibrium equation:
\begin{equation}\label{eqn:thrm}
\int_{\lambda_m}^{\lambda_M}Q_{abs}(\lambda)\cdot J^{ik}(\lambda)d\lambda- \int_{\lambda_m}^{\lambda_M}Q_{em}(\lambda)\cdot B(\lambda,T_{ik})d\lambda=0,
\end{equation}
where the equation is written for a given species of grain and 
$B(\lambda,T_{ik})$ is the black body emission. $T_{ik}$ is the temperature
of a given grain within the $ik$-th sample element, $J^{ik}$ represents 
the incoming specific intensity on the volume element, 
%
%
written, for the first iteration, as follows:
\begin{equation}\label{eqn:Iabs}
J^{ik}(\lambda)=I_{AGN}(ik,\lambda)=\frac{1}{4\pi}\cdot\frac{L(\lambda)}{4\pi r_{ik}^2}\exp[-\tau_{ik}(\lambda)].
\end{equation}
Here $\tau_{ik}(\lambda)$ is the optical depth between the central source 
and the $ik$-th sample element and $r_{ik}$ its distance.

At the second iteration the new incoming 
radiant flux on each grain of a given geometrical element $ik$ that 
will enter in the Eq. \ref{eqn:thrm} of the thermal equilibrium will be:
\begin{eqnarray}\label{eqn:Jnu}
\lefteqn{J^{ik}(\lambda)= I_{AGN}^{ik}(\lambda)+\frac{1}{4 \pi} \cdot \sum_{e=1}^{N-1}} \\ \nonumber 
\\ \nonumber 
\; & \; & \;\frac{1}{4\pi\cdot r_{ik,e}^2} \left\{ \sum_{id}^{Ndust} \left[ 4 \pi \cdot \pi a_{id}^2 Q_{id}^{em}(\lambda) B(\lambda,T_{e,id})+ \right. \right. \nonumber  \\ \nonumber  
\; & \; & \;\left.\left. 4\pi \cdot \pi a_{id}^2 Q_{id}^{sca}(\lambda) J_{tot}^e(\lambda)\right] e^{-\tau_{ik,e}(\lambda)} \right\} \nonumber  \\ \nonumber  
\end{eqnarray}
where $e$ indicates the generic volume element and $N$ is their total number.
$J_{tot}^e(\lambda)$ is the total 
{\it specific intensity} hitting the $e$--element and coming from both 
the AGN and the dust from the rest of the torus (scattering plus thermal) 
as computed from the previous step. 

These calculations are iteratively 
repeated until the difference in the temperature value between two steps 
falls below one Kelvin for each grain species in each volume element. 
If such small variations do not have big 
influence on the final result when dealing with relatively low--density 
environments, it is important to achieve a good convergence when dealing 
with high optical depth tori, to avoid losses of energy due to the numerical 
method adopted. 

\begin{figure*}
\centering
\begin{tabular}{l r}
\includegraphics[height=0.45\textwidth]{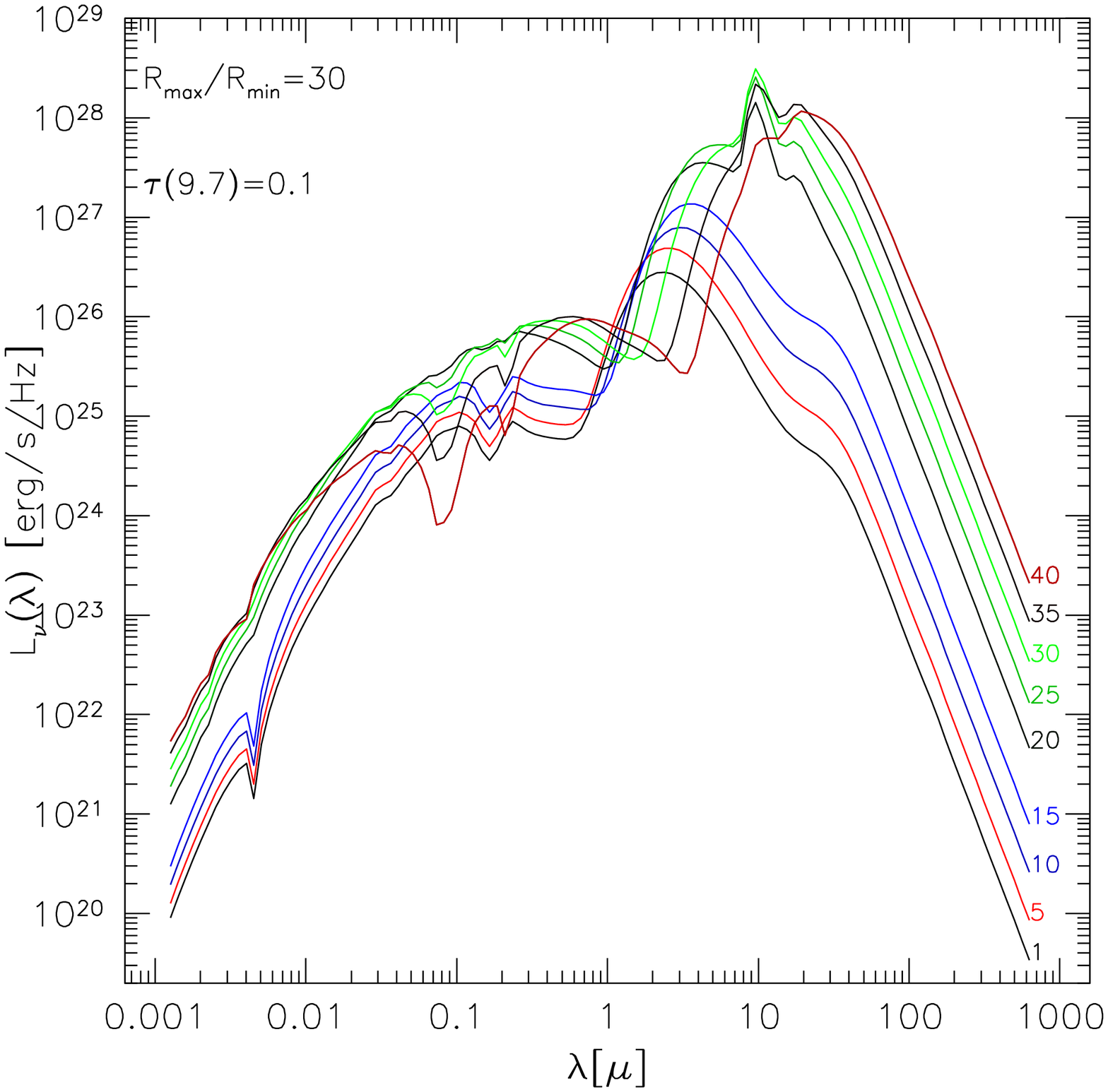}&
\includegraphics[height=0.45\textwidth]{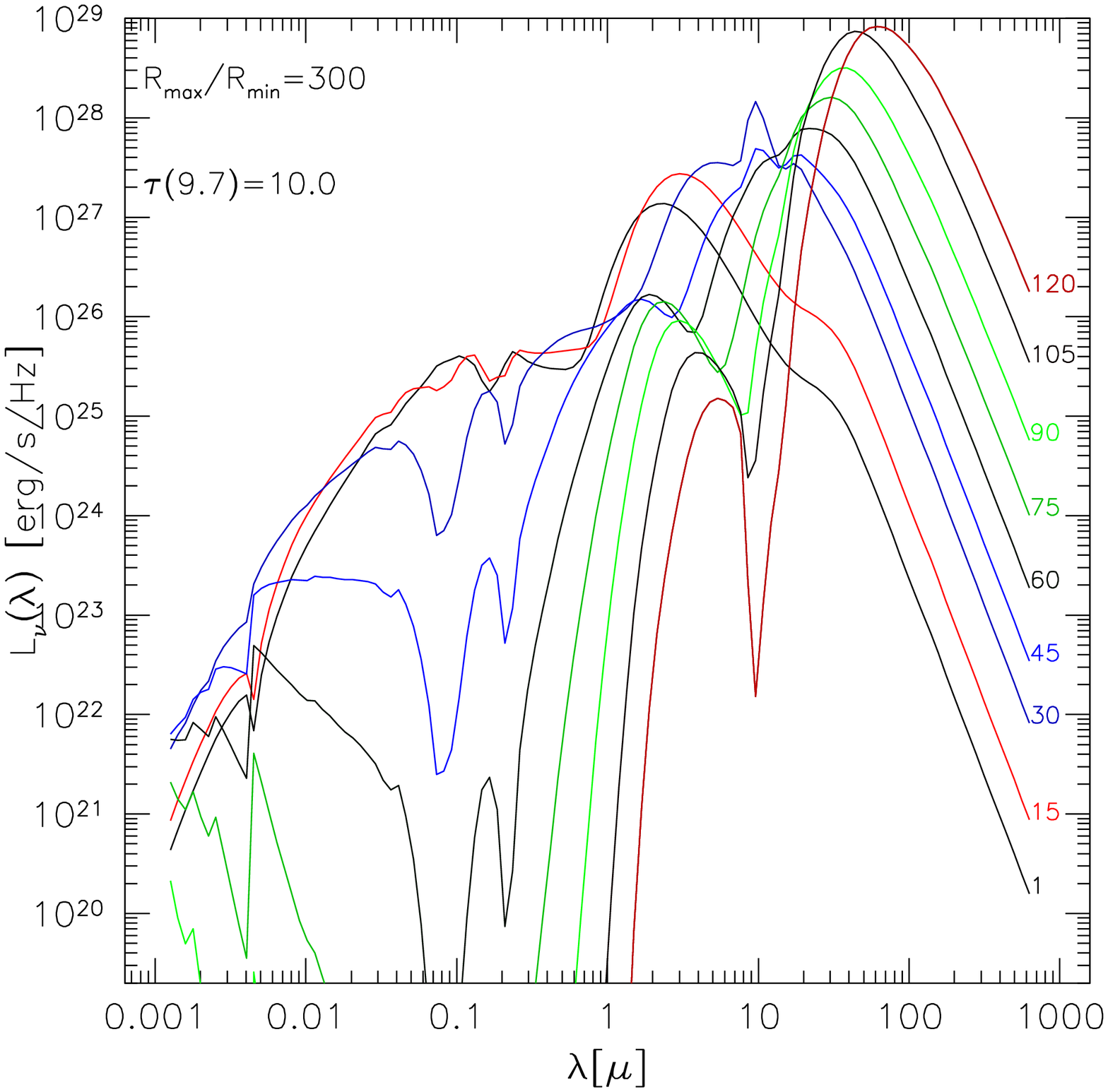}\\
\end{tabular}
\caption{Emission spectra by torus elements at increasing radial coordinate
for dust torus distributions illuminated by a nuclear source of $L^{AGN}=10^{46}$ erg s$^{-1}$.
Two extreme cases are shown. On the left panel a model with equatorial 
optical depth $\tau(9.7)=0.1$ and an outer to inner radii ratio of $30$ is
shown. The closest element to the central source is labelled $1$, 
the most distant $40$ (the different lines correspond to physical locations of 
1.36, 1.91, 2.92, 4.50, 6.8, 10.4, 16.0, 24.4 and 37.4 pc). 
On the right panel the equatorial optical depth is taken to be $\tau(9.7)=10.0$ 
and the outer-inner ratio is $300$.}
\label{fig:elements}
\end{figure*}
The conservation of energy is always verified within 1 per cent for typical
solutions and up to $10$ per cent for solutions with very high optical depth and
non-constant dust density. 
These discrepancies 
are due to some 
approximations in the maximum temperature of the grains, but do not affect
any of our conclusions.

In models with high dust densities where thermal emission strongly 
contributes to the self heating of the dust, the smaller grains 
can reach and even slightly exceed the sublimation temperature. 
In order to avoid further re-calculation of the optical depth and 
density profiles -- that would result in a strong increasing of the
computation time -- temperatures which are too high are set equal 
to the sublimation value. 
If this also happens for the $0.05$ \mums grains, then the inner 
radius is moved further out 
with respect to the value computed by means of Eq. \ref{eqn:rmin}. 

Various model solutions are reported in Fig. \ref{fig:elements}, where 
dust emission 
from elements at different distances from the central source are shown. 
The model solution on the left is built with a total opening angle of the 
torus of $140^\circ$, a ratio $R_{max}/R_{min}=30$ and a constant density 
profile with equatorial optical depth $\tau_{eq}(9.7)=0.1$. 
The minimum radius turns out to be $1.30$ pc for graphite and $6.27$ pc 
for silicate grains. 
The contribution to emission from the 
silicate grains, which is easily recognizable from the feature in 
emission at $9.7$ \mum, only appears after the element number $15$. 
The scattering component of radiation is clearly visible shortwards of 
$\lambda\sim 1$ \mum.

The model shown on the right side of the panel has the same opening 
angle for the torus, an outer-to-inner ratio of $300$, constant density 
profile and $\tau_{eq}(9.7)=10.0$. In this case the minimum radius is 
the same ($1.3$ pc) for graphite, but it is lower for silicate grains 
($3.94$ pc) whose emission appears after the element number $30$ over 
a total of $120$ radial subdivisions.

\subsection{Details on the computation of dust density and optical depth}
\label{sec:compute}

The model needs as a starting point the value of the equatorial 
optical depth at a given reference wavelength. Starting from this input 
condition we calculate the gas and dust densities and the number of the 
different dust grains within each sample element. 
With reference to Eq. \ref{eqn:rho}, providing the spatial variation of 
density through the $\beta$ and $\gamma$ parameters,
the problem is to find the $\alpha$ value yielding an
equatorial optical depth $\tau_{eq}(\lambda_{ref})$. The expression for the
latter is:
\begin{eqnarray}\label{eqn:taueq}
\lefteqn{\tau_{eq}(\lambda_{ref})=\int_{R_{min}}^{R_{max}} \left\{ \sum_{id=1}^{Ndust}\left[Q_{id}^{A} (\lambda_{ref})+Q_{id}^{S}(\lambda_{ref})\right] \right. }  \qquad \qquad\\ \nonumber 
\\ \nonumber
\; & \; & \; \left. a^2_{id} \pi \cdot Nd_{id}\cdot \rho_H(r,\theta)dr \right\}
\end{eqnarray}
where $Nd_{id}$ is the numeric density of the $id$-th grain, normalized 
to that of hydrogen. Taking this into account and the difference in the 
sublimation temperatures of graphite and silicate and the discontinuity 
in the chemical composition of dust along the radial coordinate, the 
density is a step function that can be written as:
\begin{equation}\label{eqn:dustrho}
\rho_{id}^{dust}(r,\theta)=\left\{
\begin{array}{lr}
0                             & r < R_{Sil} \\
\rho_{id}^{dust}(r,\theta)    & r > R_{Sil}
\end{array}
\right.
\end{equation}
where the indices $id$ refer only to the silicate component. An analogous
condition applies to the graphite grain distribution and their corresponding 
$id$ indices.
Substituting the $\rho$ from Eq. \ref{eqn:rho} in Eq. \ref{eqn:taueq} 
and solving for $\alpha$ we obtain the correct value of density and number 
of dust grains at each point of the grid.

The most general expression for the optical depth between any two points 
$P_1(r_1,\theta_1,\varphi_1)$ and $P_2(r_2,\theta_2,\varphi_2)$, 
located within the torus is:
\begin{eqnarray}\label{eqn:taugen}
\lefteqn{\tau_{1,2}(\lambda)=\int_{r_{1}}^{r_2} \int_{\theta_1}^{\theta_2} \int_{\varphi_1}^{\varphi_2} \sum_{id=1}^{Ndust}\left[Q^A_{id}(\lambda)+Q^S_{id}(\lambda) \right] \times} \\ \nonumber
\\ \nonumber
\; & \; & \; \qquad \pi a_{id}^2\; Nd_{id}\cdot \rho_{id}^{dust}(r,\theta) \; d\varphi \; d\theta \; dr .
\end{eqnarray}
The optical depth is calculated as the path--integral of the {\it density
function} $f(r,\theta,\varphi)=\alpha\cdot r^\beta \cdot e^{-\gamma \times |cos(\theta)|}$ 
along the straight line between the two points.
Considering the conditions \ref{eqn:dustrho} in \ref{eqn:taugen} 
automatically accounts for the fact that there are inner 
regions free of silicates due to their lower sublimation temperature 
and regions completely dust--free.

\subsection{Calculation of the emitted spectrum}

Adding the emission from all torus elements to that of
the central source, attenuated by the correct value of 
optical depth when needed, yields the total emitted spectrum.
Here it is computed for $10$ different inclination values of
line-of-sight, spanning from $0$ to $90^\circ$ with respect to 
the equatorial plane.

\begin{figure}
\centering
\rotatebox{-90}{
\includegraphics[height=.5\textwidth]{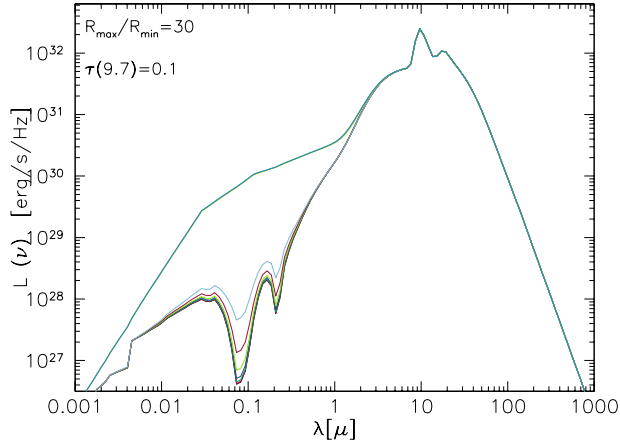}}
\caption{Emission spectra obtained for $10$ different line-of-sight 
inclinations from $0^\circ$ to $90^\circ$ at regular steps of $10^\circ$. 
Here the torus amplitude angle is $140^\circ$, the ratio between the 
outer and inner radii is $R_{max}/R_{min}=30$, the density is constant 
and the equatorial optical depth is $\tau(9.7)=0.1$.}
\label{fig:cmodel1}
\end{figure}

\begin{figure}
\centering
\rotatebox{-90}{
\includegraphics[height=.5\textwidth]{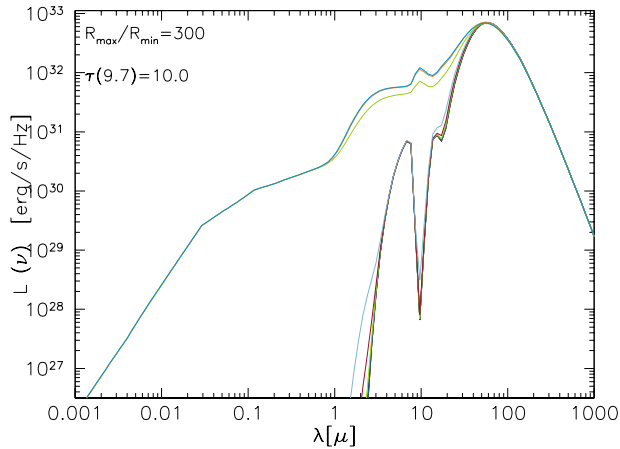}}
\caption{Emission spectra as a function of wavelength for $10$ different 
lines-of-sight (see also Fig. \ref{fig:cmodel1}) for a geometrical 
configuration with $\tau(9.7)=10.0$, $R_{max}/R_{min}=300$, torus
opening angle $\Theta=140^\circ$ and constant density profile.}
\label{fig:cmodel2}
\end{figure}

In Figs. \ref{fig:cmodel1} and \ref{fig:cmodel2} we show two emission 
models with different lines-of-sight and 
different physical and geometrical parameters. The first model is meant 
to represent a geometry in which the amplitude of the torus is $140^\circ$, 
the ratio between external and internal radii is $R_{max}/R_{min}=30$, 
the density is constant and the optical depth is $\tau(9.7)=0.1$. 
Such a configuration is obtained with a hydrogen column density of 
$9.0 \times 10^{21}$ cm$^2$ and corresponds to an optical extinction 
of $A_V=2.3$. 
Note that the shape of the SED remains quite constant with varying angle, 
as long as the central source is seen directly.
A similar behaviour is 
found when both the optical depth and the torus size
increase. The emission pattern of Fig. \ref{fig:cmodel2} has a radial
ratio of $300$ and the equatorial optical is $\tau(9.7)=10.0$. 

\section{MODEL REALISATIONS}
\label{param}

We have computed a grid of models by varying the different parameters.
Some of their main features and a rough comparison with 
other models in literature are presented here.

\subsection{The grid of model solutions}
\label{sec:modgrid}

We obtained a set of models spanning the widest range of physical 
and geometrical characteristics of dusty tori discussed in the literature. 
The models have varying radial ratios 
$R_{max}/R_{min}=30,\; 100,\;300$, corresponding to radial sizes
from about $40$ to about $400$ pc (for an accretion luminosity 
of $10^{46}$ erg s$^{-1}$), while the torus full opening angles were taken to be $60^\circ$, 
$100^\circ$ and $140^\circ$, corresponding to sizes of the torus 
between a minimum of $\sim 20$ to a maximum of $\sim 370$ pc. 
The equatorial optical depth was always measured at $\lambda =9.7$ \mums 
and the models have been computed at values of $\tau(9.7)=0.1$, $0.3$, $0.6$, 
$1.0$, $2.0$, $3.0$, $6.0$ and $10.0$, translating into values of 
absorption in the $V$ band from few to few hundreds magnitudes, 
depending also on the $R_{max}/R_{min}$ ratio. The parameters used 
to define the law for the spatial behaviour of density were 
$\beta=0, -0.5$ and $-1.0$ and $\gamma =0$ and $6$ (see Eq. \ref{eqn:rho} 
for the definitions). In order to have a full exploration of the parameter 
space, we also considered models with density increasing towards the outer parts 
of the torus, obtained by setting $\beta=0.5$. 
In Table \ref{tab:mod_grid} we summarize the main features of the model grid.

\begin{table}
\centering
\begin{tabular}{l c l}
\hline
Parameter	       &  & Adopted values  \\
\hline
R$_{Max}$/R$_{min}$    &  & $30$, $100$, $300$ \\
$\tau_{eq}(9.7\mu m)$  &  & $0.1$, $0.3$, $0.6$, $1.0$, $2.0$, $3.0$, $6.0$, $10.0$ \\
$\beta$ 	       &  & $0.5$, $0.0$, $-0.5$, $-1.0$ \\
$\gamma$	       &  & $0.0$, $6.0$ \\
$\Theta$	       &  & $60^\circ$, $100^\circ$, $140^\circ$ \\
\hline
\end{tabular}
\caption{The parameters of the models of the grid.}
\label{tab:mod_grid}
\end{table}

\subsection{Main features of the models}

\subsubsection{The distribution of grains temperatures}

One of the most evident features of the AGN IR emission is the presence
of multiple temperature components. Especially when viewed face-on, the
whole range of temperatures can be seen, from the innermost part -- where
grains are close to the sublimation limit -- 
to the outer shells where the dust is colder, only heated by reprocessed 
IR radiation.

\begin{figure*}
\centering
\begin{tabular}{l l}
\includegraphics[height=.45\textwidth]{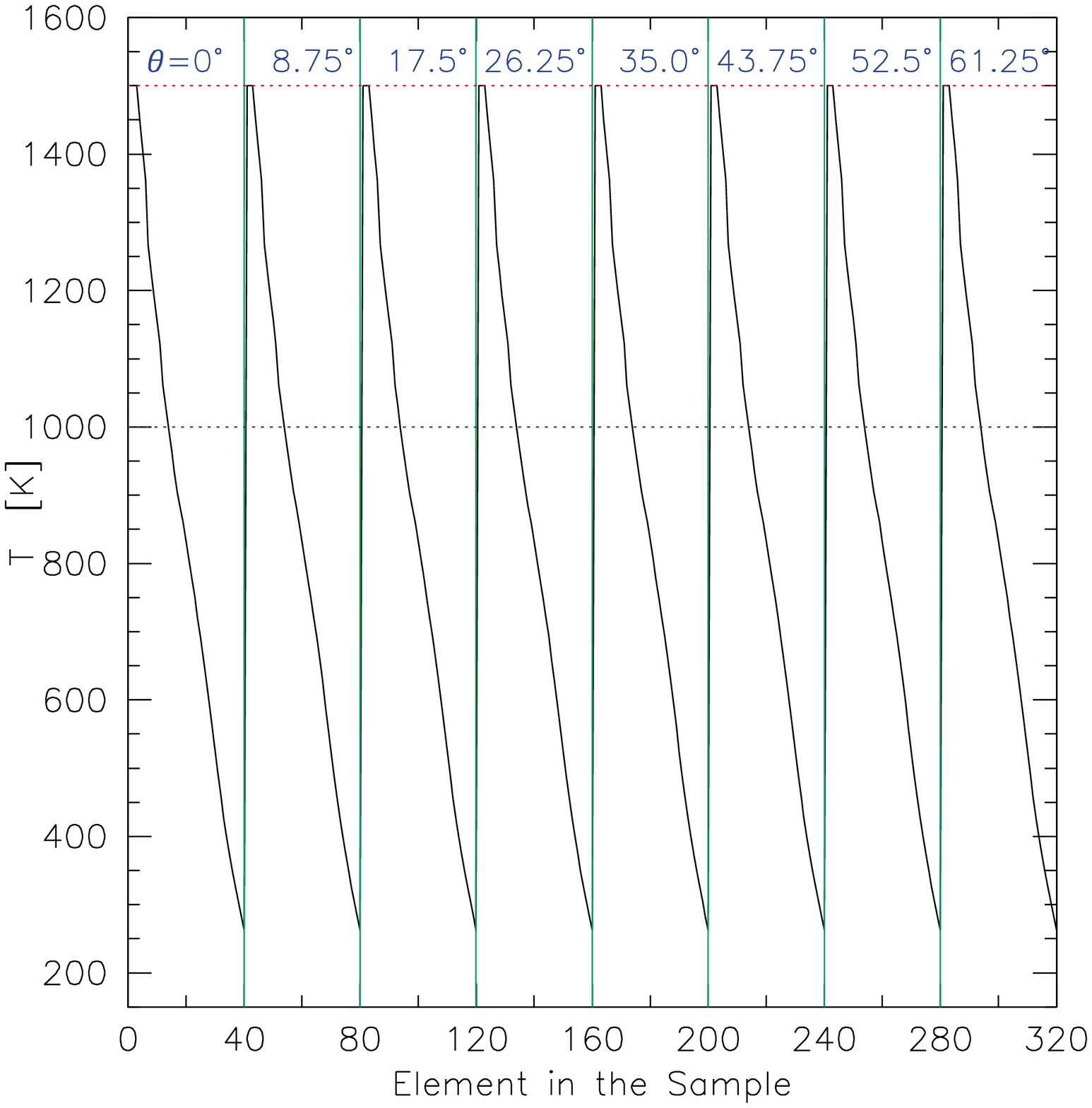} &
\includegraphics[height=.45\textwidth]{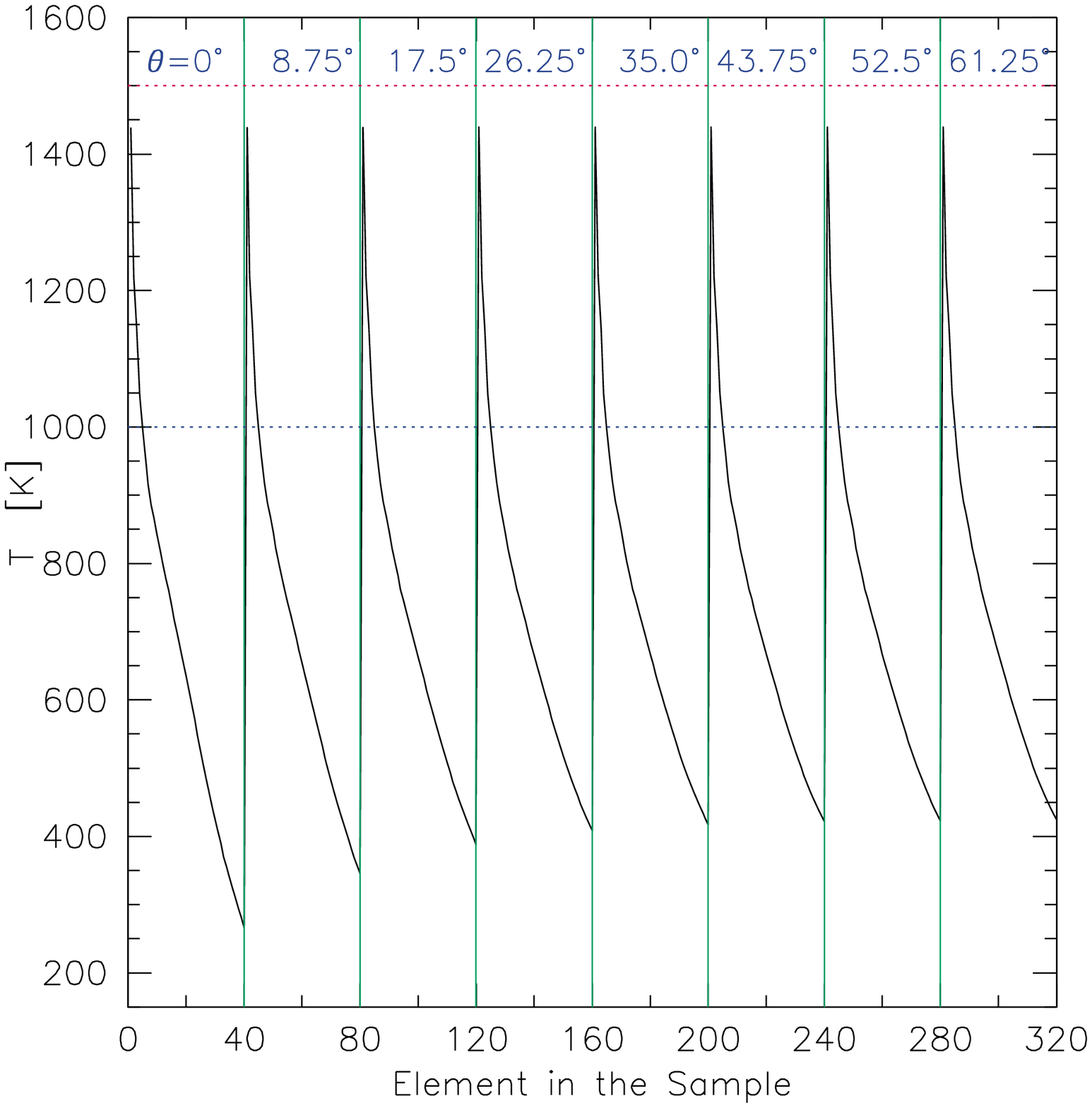} \\
\end{tabular}
\caption{Temperature profiles of graphite grains with a radius of $0.01$ \mums
as a function of the torus element. Each vertical blocks spans
the radial distance of the torus and corresponds to a different 
polar angle. On the left panel a model with constant density profile, $\tau=0.1$ 
and radii ratio of $30$ is shown. In this case the only temperature 
gradient present is the one with respect to the radius. On the right 
panel a model with the same characteristics but with also a non-constant 
density profile ($\beta=-1.0$ and $\gamma=6$). The horizontal lines 
represent the values of sublimation temperature assumed for graphite 
($1500$ K) and for silicate ($1000$ K) grains respectively.}
\label{fig:temperatures}
\end{figure*}

For models in which the spatial profile of density is kept constant with 
respect to the altitude angle $\theta$, the temperature is assumed to 
vary only along the radial coordinate. Models with non-constant density 
profile can show a temperature gradient also with respect to the altitude 
coordinate. In Fig. \ref{fig:temperatures} we show the trend of the dust
equilibrium temperature as a function of the coordinates of the sample element. 
Two extreme cases are shown: a model with $\tau(9.7)=0.1$, 
$R_{max}/R_{min}=30$ and a constant density, and a model with 
the same parameters but a density profile varying like 
$r^{-1.0}\cdot e^{-6\times |\cos(\theta)|}$ 
(in both cases a graphite grain with radius of $0.01$ \mums is considered).

In the case of a constant dust density distribution, the temperature profiles 
drop roughly linearly with the distance from the central source. 
A non-constant density profile, in our case peaking in the inner part of 
the torus (close to the central source and to the equatorial plane), 
brings some differences. First of all, the decrease of dust density 
with increasing altitude on the equatorial plane results in lower 
values of the optical depth of the elements at lower $\theta$ 
coordinate. The temperature 
in the most distant volume elements will then increase with 
increasing altitude with respect to the equatorial plane. 
Furthermore, since the dust density peaks in the innermost 
regions, the temperature gradient will be higher for those 
volume elements that are closer to the inner edge of the torus. 
Moreover the high density (hence high optical depth) values that 
are found close to the inner radius are responsible for not always
reaching the sublimation temperature:
in such configurations the central source undergoes a high 
attenuation in the first layers of dust.

\subsubsection{Dependence on the viewing angle}

As expected for a geometry with no spherical symmetry, the emission 
is highly anisotropic especially when the lines of sight that cross the 
torus and those that see the AGN directly are considered (see Figs. 
\ref{fig:cmodel1} and \ref{fig:cmodel2}). This feature is further 
highlighted if one considers a pattern in which the density is not 
constant along the altitude angle $\theta$. 

Figs. \ref{fig:cmodel1} and \ref{fig:cmodel2} show that
the SED remain quite constant even with varying 
line-of-sight angle, until the latter intersects the dust-free zone, 
i.e. at $\Psi <90^\circ-\Theta/2$. The only part of the emission that changes 
is the one shortwards of $1$ \mums for low optical depth models, 
where the emission of the non-thermal central source is directly 
seen in the type-1 scheme configuration.
In high optical depth systems this behaviour is maintained, 
but the drastic difference between type-2 and type-1 concerns 
the emission shortwards of $30$ \mum.

If a non-constant density pattern is assumed 
the change is smoother since the emission from the central source becomes 
more visible as the torus is seen at decreasing polar angles, where the density 
is assumed to be lower with respect to the equatorial plane. Notice that 
with such density profiles and in those models with low optical depth 
values the direct view of the central source is achieved even when 
$\Psi >90^\circ- \Theta/2$ since the optical depth at higher altitudes 
becomes small enough. Then a substantial contribution of the accretion 
disc emission can be seen even through the dust cone. Configurations
with decreasing number of dust grains with increasing altitude on the 
equatorial plane can be considered to mimic the tapered torus 
geometry analysed by \cite{efstat95} since, at least for
low values 
of the equatorial optical depth [i.e. $\tau(9.7)<1.0$], the bulk of 
emission comes from those volume elements that are closer to the equatorial plane. 
The other elements of the geometrical grid will in fact provide a 
smaller contribution to the total emission, as they will contain a 
smaller number of grains.

\subsubsection{The width of IR spectrum}

\begin{figure*}
\centering
\begin{tabular}{c c}
\includegraphics[height=.45\textwidth]{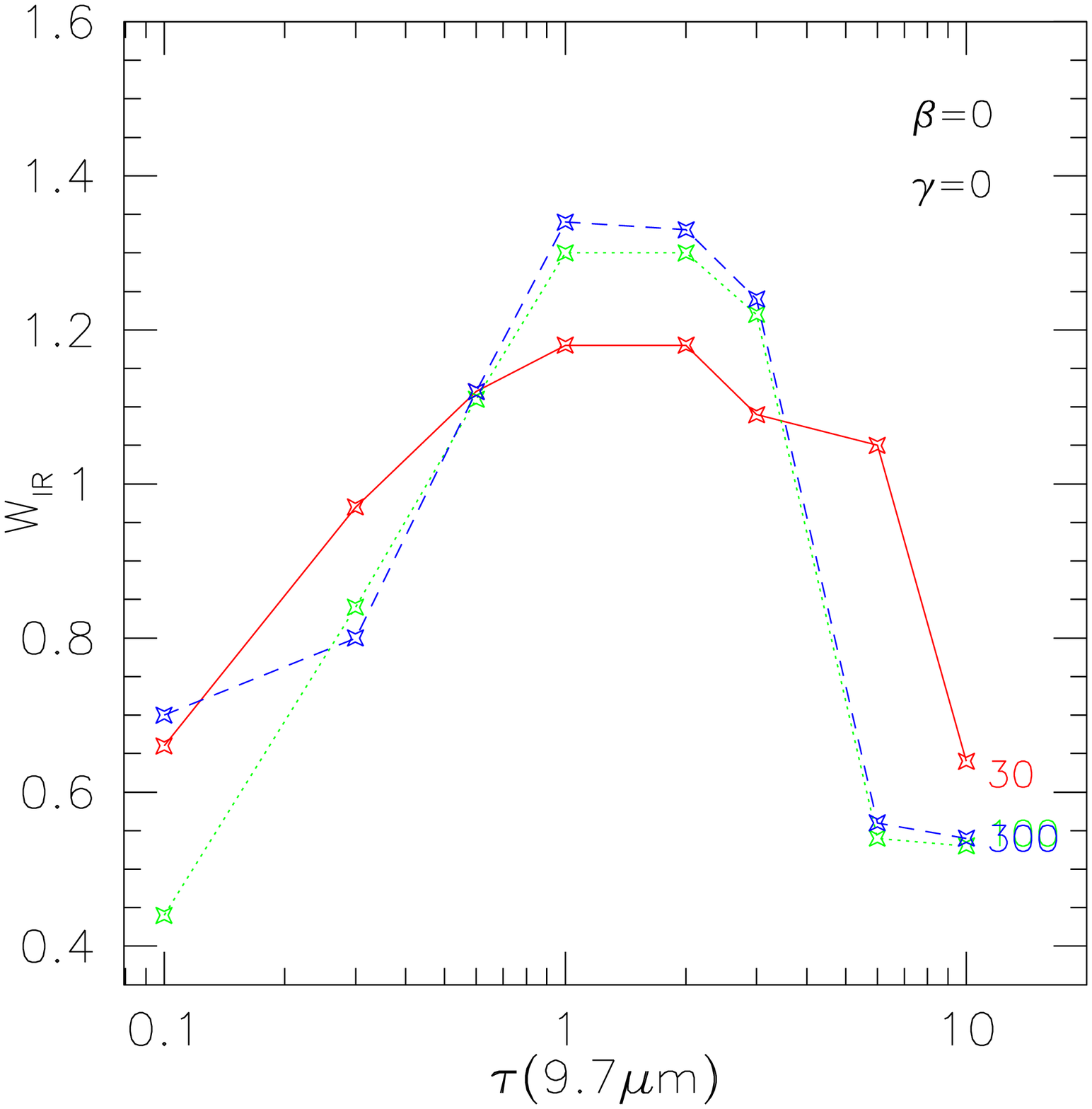} &
\includegraphics[height=.45\textwidth]{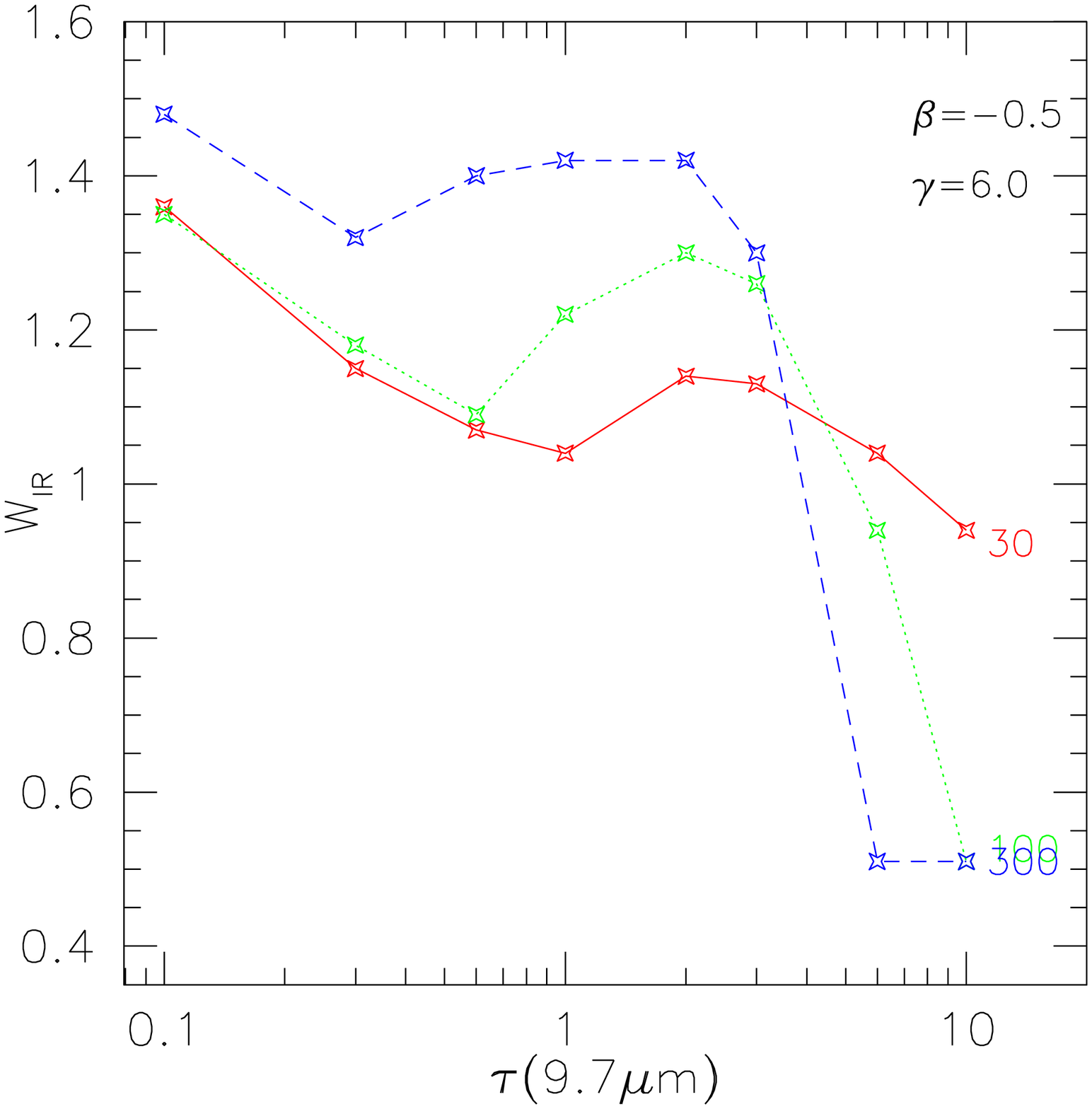} \\
\end{tabular}
\caption{The trend of the IR bump width is shown as a function of the optical depth and for various radii ratios. The parameter $W_{IR}$ is defined as the logarithmic interval in wavelength, were the power emitted is more than one third of the maximum. On the left-side panel models with homegeneous density distributions are shown, while on the right panel a model with density parameters $\gamma=6.0$ and $\beta=-0.5$ is plotted.}
\label{fig:bump}
\end{figure*}

One of the most important features of a dusty torus model is the capability 
of reproducing the observed width of the IR spectrum due to multi-temperature  
emission. Both \cite{pierkrolik} and \cite{granato94} use for its measure the 
logarithmic wavelength interval in which the power $\lambda F_\lambda$ emitted 
in the IR is more than one third of the peak value ($W_{IR}$). When measured on 
a black body this parameter has a value of $\sim 0.7$ while, for example, in the 
observed spectra used by \cite{granato94} its value is always larger than $1.3$. 
In Fig. \ref{fig:bump} we report the trend of the bump width as a function of 
optical depth for different torus sizes, as viewed along the torus equatorial 
plane. This parameter depends both on the optical depth value and on the torus 
size: increasing the values of the optical depth will tend to generally shrink 
the bump since the inner, hotter, parts will become less visible. At rising values 
of the radii ratio the bump will tend to broaden, since larger radii imply higher 
amounts of colder dust.

There are a few considerations to be made on the interpretation of the $W_{IR}$ trend in Fig. \ref{fig:bump}. 
The maximum of emission for low values of optical depth falls right on 
top of the silicate feature which in these cases can display a very 
intense emission. This will in turn give a low value for the bump width 
as defined by \cite{pierkrolik}. Furthermore, very high values of optical 
depth will cause the silicate feature to appear in strong absorption 
especially at equatorial optical depth higher than $6$, resulting,
again, in a low width for the IR bump. 

A density profile peaking towards the innermost regions of the torus is less 
affected by this behaviour. This is mainly due to the fact that in the latter 
case the silicate feature is less prominent both in emission and in absorption, 
compared to a constant dust distribution configuration.

\subsubsection{The peak of the IR emission}

\begin{figure*}
\centering
\begin{tabular}{l l}
\includegraphics[height=.45\textwidth]{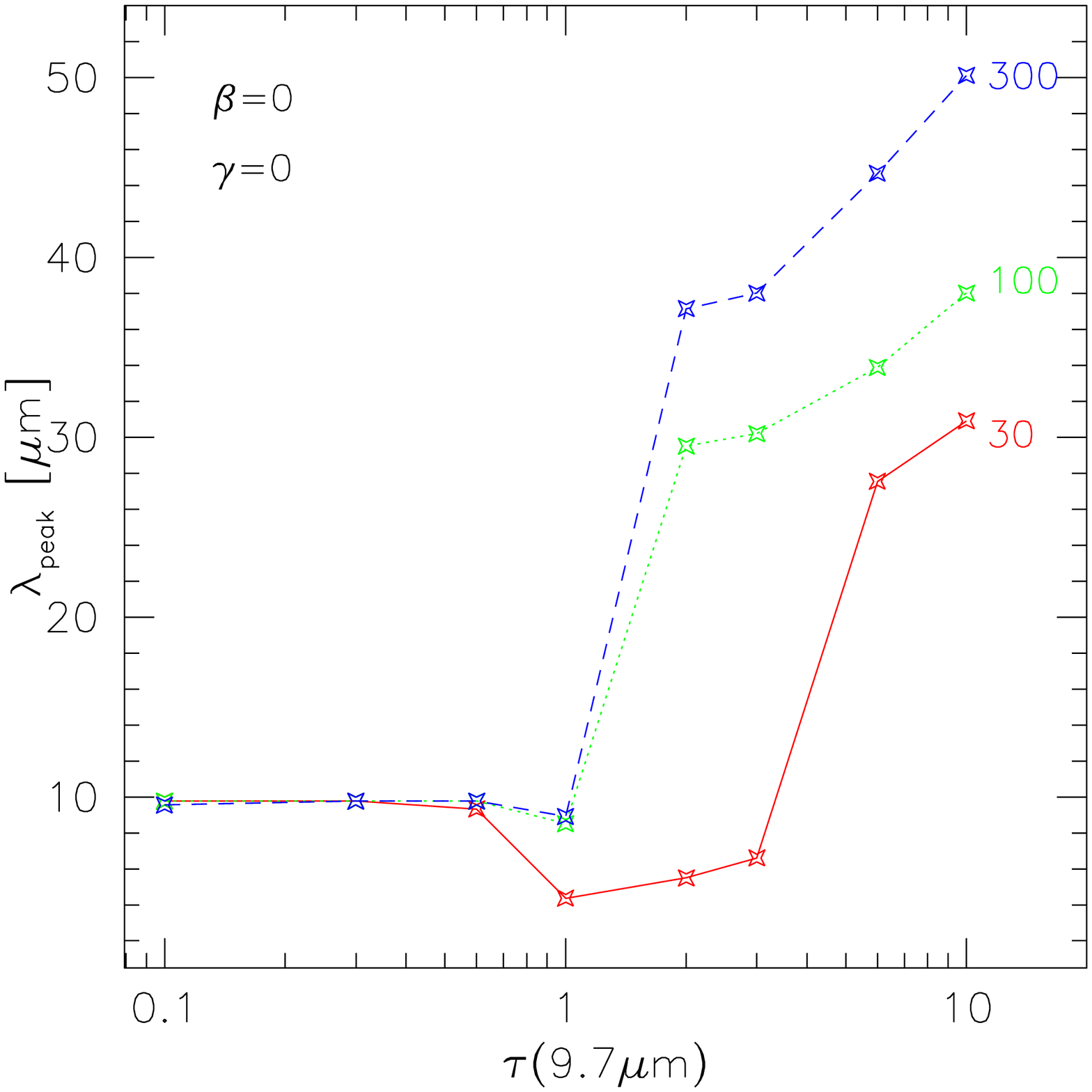} &
\includegraphics[height=.45\textwidth]{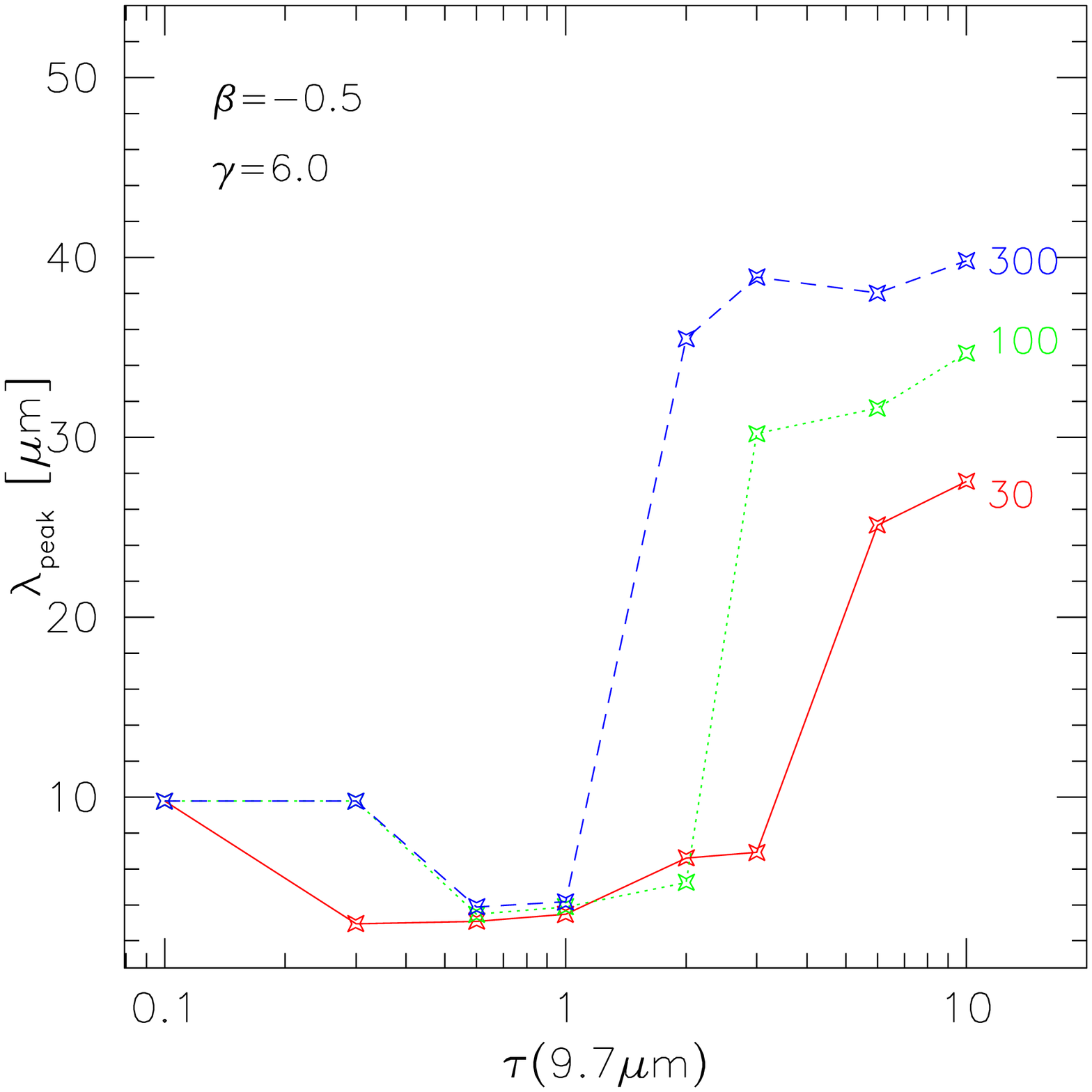} \\
\end{tabular}
\caption{Dependence of the wavelength of the peak IR emission (expressed in $\lambda F_\lambda$) on the optical depth and torus size for a constant (left panel) and a non-constant (right panel) density distribution in the torus, calculated for a type-2 viewing angle. The constant value of $\sim 10$ \mums for small values of optical depth is due to the presence of the intense $9.7$ \mums silicate emission dominating the spectrum. Here the dependences on the optical depth and on the torus size are comparable. No significant differences are found for the non-constant density model.}
\label{fig:lpeak}
\end{figure*}

Another important parameter that can be used to characterize a model solution is $\lambda_{max}$, the wavelength at which the IR emission -- expressed as $\lambda F(\lambda)$ -- peaks. In Fig. \ref{fig:lpeak} this parameter is shown against the optical depth and for the three torus sizes for which the model has been computed. Only type-2 (equatorial view) emission is shown here. The peak wavelength ranges from $9.7$ \mum, corresponding to the presence of a strong silicate emission in models with low optical depth, to about $60$ \mums obtained in the models with the largest outer radii and optical depth values. 

A non-constant dust density profile results in a hotter emission, 
with the emission peaking at smaller wavelengths. This occurs because 
lower densities -- and hence lower optical depths -- on the outer 
regions of the torus make the inner, hotter, regions more visible to 
the observer, increasing this way the average observed temperature.

\subsubsection{The $9.7$ \mums silicate feature}

The presence of the absorption feature at $9.7$ \mums in the spectra of Seyfert 2 galaxies has been used in the literature as a proof for the presence of silicate grains. At the same time, IR 
observations often showed rather featureless spectra for Seyfert 1 galaxies, which brought into question
the reliability of the Unified Model: the characteristics of the silicate emission 
at this wavelength has become a crucial test of the validity of models. 
\cite{nenkova02} suggested that a clumpy distribution of dust may provide 
an explanation for the lack of the silicate feature in type-1 sources, 
while \cite{dullemond-bemmel} claim that the same result can be achieved 
in smooth tori. 
The depth of the same feature in type-2 sources, however, is shallower in clumpy tori.

\begin{figure*}
\centering
\begin{tabular}{l l}
\includegraphics[height=.45\textwidth]{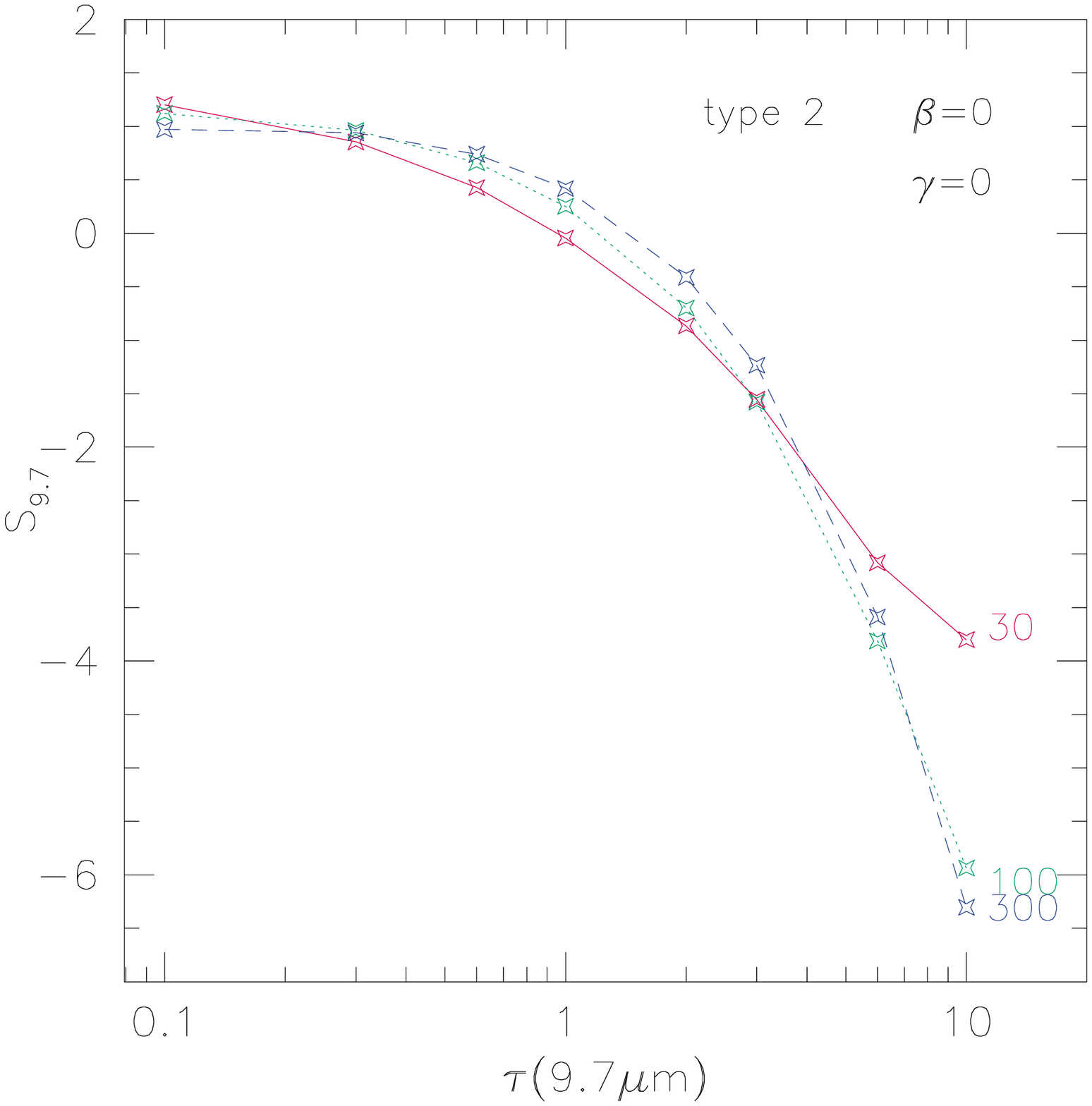} &
\includegraphics[height=.45\textwidth]{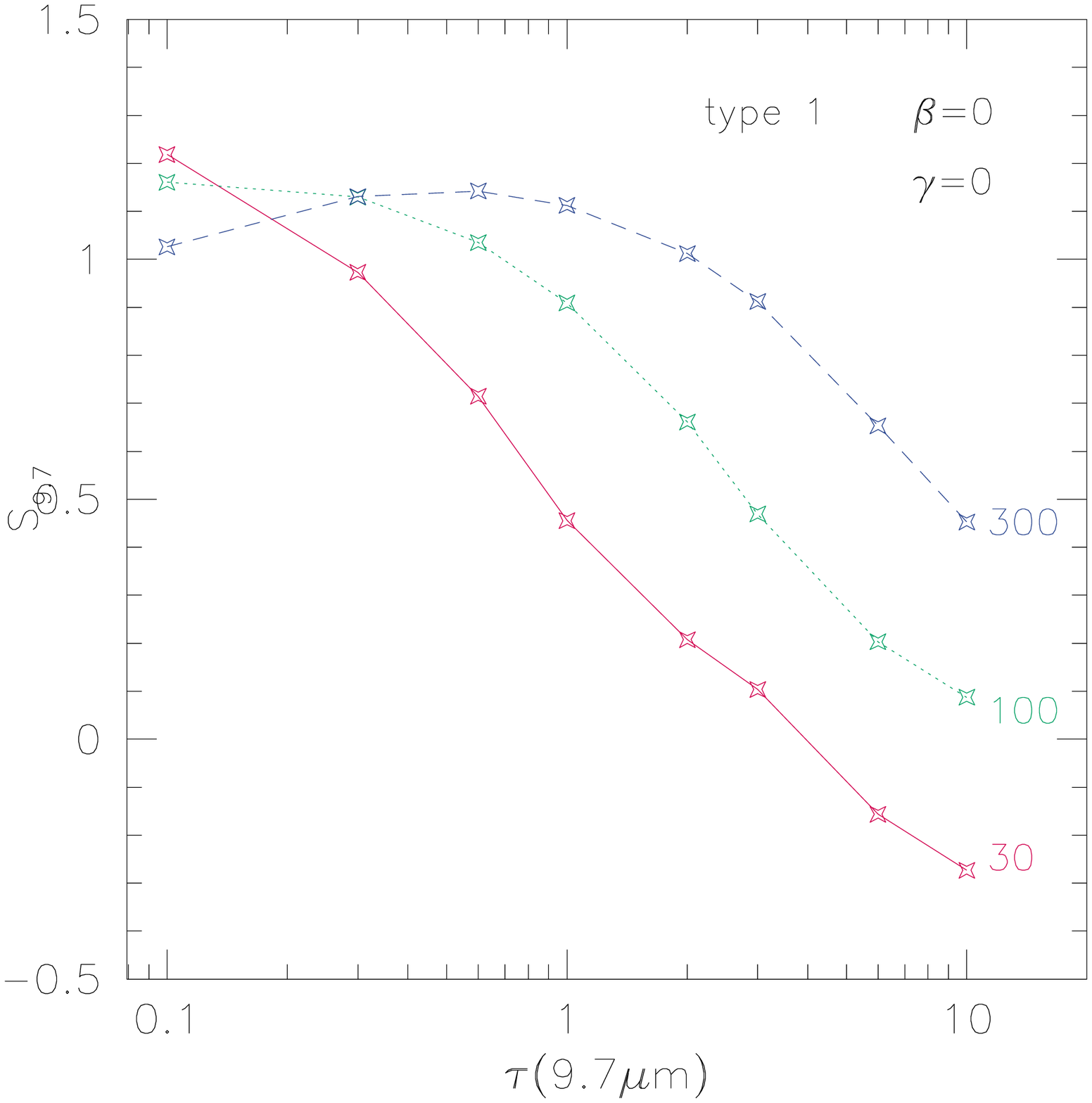} \\
\includegraphics[height=.45\textwidth]{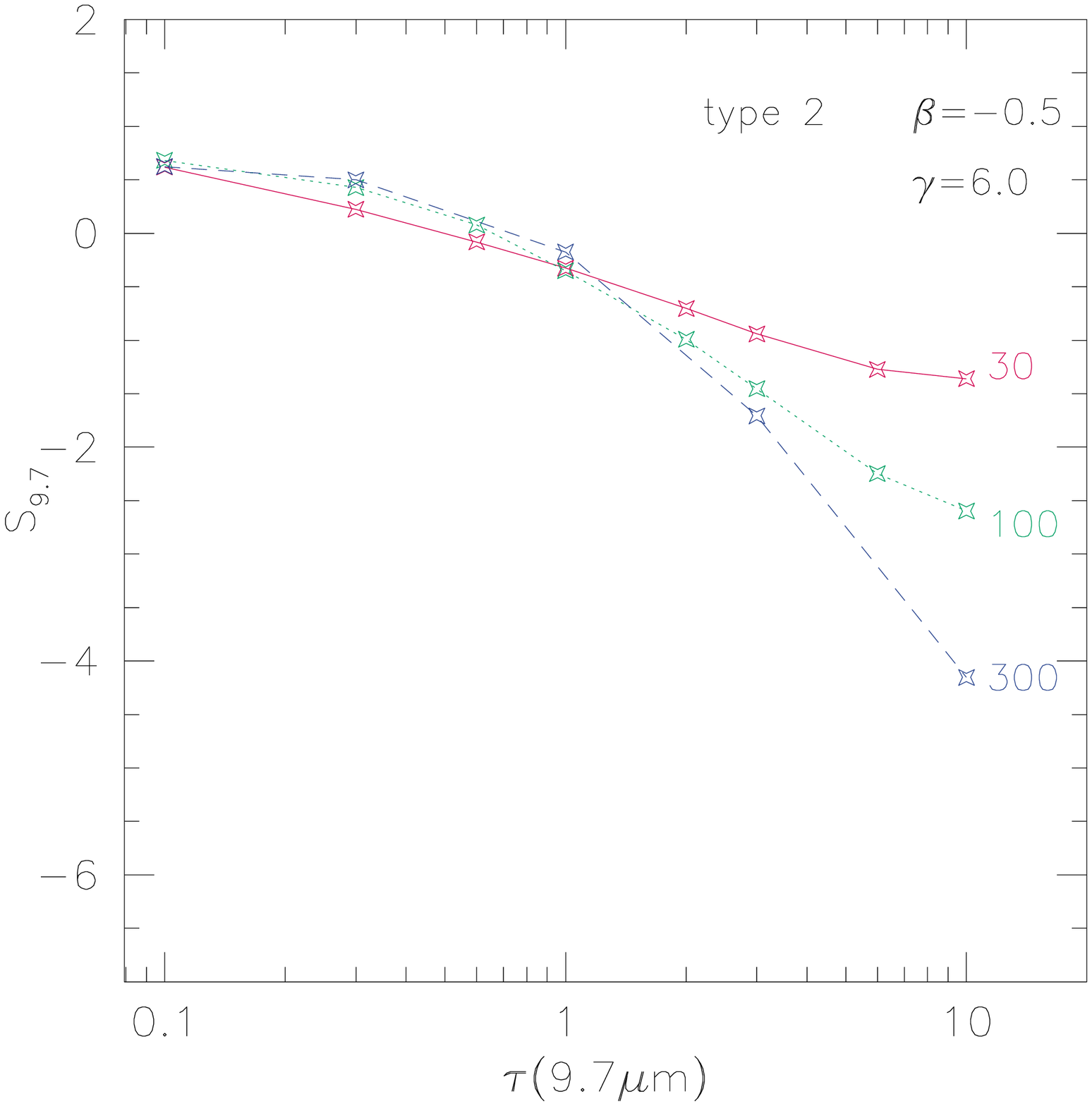} &
\includegraphics[height=.45\textwidth]{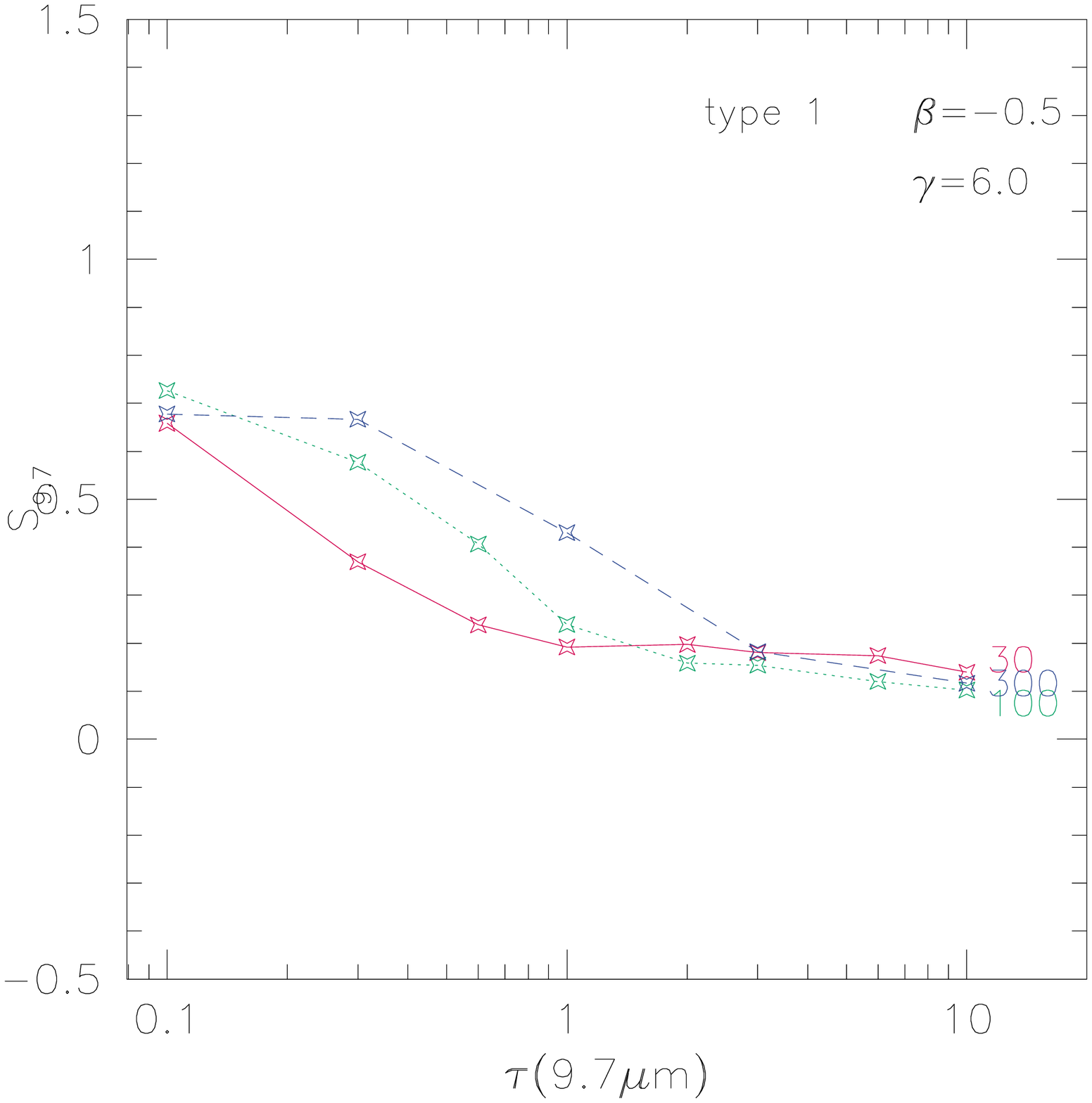} \\
\end{tabular}
\caption{The variation of the $9.7$ \mums feature intensity $S_{9.7}$ as a function of the optical 
depth for three torus sizes. In the two panels at the top, a torus with a constant density distribution is considered, 
while at the bottom a model with a density gradient is shown ($\beta=-0.5$ and $\gamma=6.0$). 
A type-2 line-of-sight ($\Psi=90^\circ$) is considered on the left-side plots: here we see that 
the strongest dependence is on the optical depth rather than the torus dimensions. On the right-hand 
panels the type-1 view ($\Psi=0^\circ$) is shown. The variation of the $S_{9.7}$ parameter 
is smoother in this case and the most extreme variations are encountered in models with the 
smaller sizes. Models with the density peaking in the innermost regions show a general smaller variation.}
\label{fig:silfeat1}
\end{figure*}

Only very recently \cite{siebenmorgen} and \cite{leihao} have found clear observational evidences for the presence of the $9.7$ \mums silicate emission feature in type-1 quasars from spectral observations by the Spitzer Space Telescope. 

Following \cite{pierkrolik} and \cite{granato94}, we measured the amplitude of this feature by means of the dimensionless parameter $S_{9.7}$, defined as the natural logarithm of the ratio between the actual flux emitted at $9.7$ \mums and the flux value at the same wavelength interpolated between $6.8$ and $13.9$ \mum:
\begin{equation}
S_{9.7}=\ln \left[\frac{F(9.7)}{F_C(9.7)}\right]
\end{equation}
In Fig. \ref{fig:silfeat1} we report the variation of the parameter $S_{9.7}$ as a function
of the external to internal radial ratios and of the optical depth for a type-2 (left panel) and type-1 lines-of-sight (right panel). A homogeneous distribution of dust is assumed for models in the upper panels while in the lower panels the behaviour of non-constant density models is shown.

The strongest dependence of the silicate emission for type-2 viewing angles is on
the optical depth, while the torus size plays a secondary role that becomes more important when a type-1 viewpoint is assumed. In the latter case the variations are much smoother and smaller torus sizes appear to be more sensitive to changes in optical depth.
Hence, assuming a density profile that decreases towards the outer zones makes the interval of $S_{9.7}$ values more narrow with respect to a constant density profile, both in the type-2 and in the type-1 solutions. This is likely due to the fact that the silicate features in such models is more nested in high optical depth zones, and its presence is attenuated. 

Models in which the silicate feature is absent or even measured in absorption 
for a polar (type-1) view are obtained with high values of the equatorial optical depth [$\tau(9.7)$ around $10$], smaller outer-to-inner radius ratio and $\gamma =0$. The amount of absorption increases as the exponent $\beta$ decreases, while positive values yield an almost featureless spectrum. Models with decreasing density at increasing angular
distance from the equatorial plane always display the silicate feature in emission, as 
the hotter innermost regions  where the silicate feature is stronger are less extinguished 
when the torus is seen face on. 
The intensity of the Silicate emission feature predicted by our model -- thanks to the detailed description of the
inner torus boundary where the feature is mainly produced -- is moderate, and almost absent for $\tau(9.7)>1$ (see Fig. \ref{fig:silfeat1} bottom-right panel).

\subsection{Scaling properties of dust emission}

Some considerations must be made in order to enable the derivation of various physical quantities for each source from the comparison of the model expectation with the observational data. 

All models are computed assuming a luminosity of the primary power source of $L^{AGN}=10^{46}$ erg s$^{-1}$. Both the intensity of dust emission and the torus sizes are  strongly related to this value, as clearly indicated by Eq. \ref{eqn:rmin}
about the relationship between the mininum radius and the AGN luminosity. 
The problem is that of finding how the solutions of the radiative transfer 
equation depend on the primary power source luminosity and on the size of the
emitting region, and how we can
scale the spectral solutions from the assumed $L^{AGN}=10^{46}$ erg s$^{-1}$ to the
value required by the target source.

\cite{rr80} notes that the radiative transfer problem has scaling 
properties that allow a reduction of the number of independent input parameters. 
These properties were first derived for spherical-shell geometries, but 
\cite{ivezic97} showed that they are a general property of dust emission in any 
arbitrary geometry. 
They showed in particular that main pivotal parameter in the dust
emission modelling is the overall optical depth. 
So, as long as the optical depth remains the same, 
the size of the system can be scaled 
up or down arbitrarily, without affecting the radiative transfer solution.
Obviously, the scale size of the torus distribution determines the
value of the dust mass.

Then, assuming a given value for the equatorial optical depth, if we change
the source luminosity and size, we have the following scaling
relations:
\begin{equation}
\tau \propto \rho \cdot R \Rightarrow \; \; \rho \propto R^{-1} \Rightarrow \; \; \frac{M}{R^3} \propto R^{-1} \Rightarrow \; \; M\propto R^2
\end{equation} 
where $R$ is the torus radius and $M$ is the total dust mass. 
On the other hand, the torus luminosity is proportional to the grain 
number density and the grain emissivity $\epsilon$, that is 
$L \propto {\rho}\epsilon / m \propto M$, where $m$ is the average grain
mass.
Altogether, total dust mass and torus luminosity scale in the same way, 
like the square of the linear size.

The mass of dust for a geometrical configuration of the torus is simply 
calculated as the sum of the mass of all the grains contained in each 
volume element, summed over all the elements of the grid. 
The mass of a dust grain is calculated from the typical 
density of the corresponding grain specie, by means of: 
\begin{equation}\label{eqn:dustmass}
m_i=\frac{4}{3}\pi\cdot a_i^3\cdot \rho
\end{equation}
where the $i$ index is meant to represent grains with given size $a_i$
and $\rho$ is the mass density of the specie, 
which was assumed to be $2.26$ and $3.50$ gr/cm$^3$ for 
graphite and silicate respectively (see e.g. \citealt{weingartner01}).

To summarize, the parameters of the torus model include the torus opening 
angle $\Theta$, the ratio $R_{max}/R_{min}$ of the outer-to-inner radii, the coefficients $\beta$ 
and $\gamma$ that enter the density function, 
and the equatorial optical depth measured at $9.7$ \mum. 
Other parameters, like for example those describing the distribution 
function for the sizes
of dust grains or their maximum and minimum values, were kept constant in 
order to avoid further complicating the problem.

A final quantity to consider is the torus viewing angle $\Psi$: 
apart from specific 
cases, we will henceforth assume that it is either $\Psi=0 ^\circ$ for a polar view 
needed to represent type-1 AGN, or $\Psi=90 ^\circ$ for an equatorial view
needed for type-2 AGN.

\section{Modeling the SED of Active Galactic Nuclei}
\label{fit}

In the present Section we exploit our AGN IR model to study the observed 
broad band SEDs of a variety of mostly nearby objects, known to host active 
nuclei, selected from the {\it NASA/IPAC Extragalactic Database} (NED). We 
confine the study to sources with numerous (at least ten independent) 
available photometric datapoints in order to ensure a good sampling of their SEDs.

A particularly interesting class of active galaxies are the 
Ultra-Luminous and Hyper-Luminous InfraRed Galaxies (ULIRGs and HyLIRGs, respectively).
The studies of ULIRGs by {\it IRAS}
and later by {\it ISO} often revealed the simultaneous presence of starburst 
and significant AGN dust components, whose respective contributions range from 
a small fraction up to the totality of the IR SED  (e.g. \citealt{genzel98}; 
\citealt{verma02}; \citealt{berta03}; \citealt{prouton04}). 
The IR emission of a starburst typically peaks at wavelengths in the range between $\sim50$
and $100$ \mums (see e.g. \citealt{mrr84}; \citealt{soifer84}), while the peak of
AGN emission occurs at shorter wavelengths, somewhere between $3$ and $40$ \mum.
Hence, given the fact that AGN and starburst phenomena often happen concomitantly, 
and that in nearby AGN the host galaxy component can make an important contribution 
to the optical and near-IR spectrum, all three components have been included in our 
spectral modeling. 
For the cold dust component, the major contributor to the emission at wavelengths
longer than $\sim 50$ \mum, we chose to use starburst templates (see e.g. \citealt{farrah03}). 
More precisely, M82 represents a typical starburst IR emission while Arp 220 is
adopted as representative of a very extinguished starburst.
Additionally, NGC 1482, NGC 4102, NGC 5253 and NGC 7714, 
are used in order to widen the choice of spectral shapes. The main differences 
between these templates are the intensity of the PAH features (between $\sim 6$ 
and $\sim 15$ \mum), the depth of the silicate feature at $9.7$ \mums and the 
width, intensity and peak wavelength of the IR bump.

A more exhaustive approach would require to provide also a physical 
description of the starburst component (e.g. \citealt{efstat05}), 
which however is far beyond the scope of this work.
Our simplified approach of using instead observational templates of local 
starbursts is in line with the current work aimed at
modelling the AGN dusty tori and deriving values for the best-fit
parameters from comparison with the data: as explained below, 
starburst templates are added only when a torus model fails to 
provide an acceptable description of the observational SED. Concerning the starburst contribution, it is mainly determined by 
the far-IR datapoints, predominantly from {\it IRAS} 60 and 100 \mums measurements,
which are prone to considerable uncertainities. For this component
we limit our analysis to just an inference of the bolometric IR luminosity. Finally, the stellar component, when needed, is
modelled using optical model spectra of ULIRGs in the wavelength range from $0.3$ to
$\sim 5$ \mums, as discussed by \citealt{berta03} and Fritz et al. (in preparation).
Note that we left free the relative normalization of this optical/near-IR 
component and the far-IR spectrum of the host galaxy, given the extremely
complex physical relation of the two.

\begin{figure*}
\centering
\rotatebox{270}{
\includegraphics[height=0.8\textwidth]{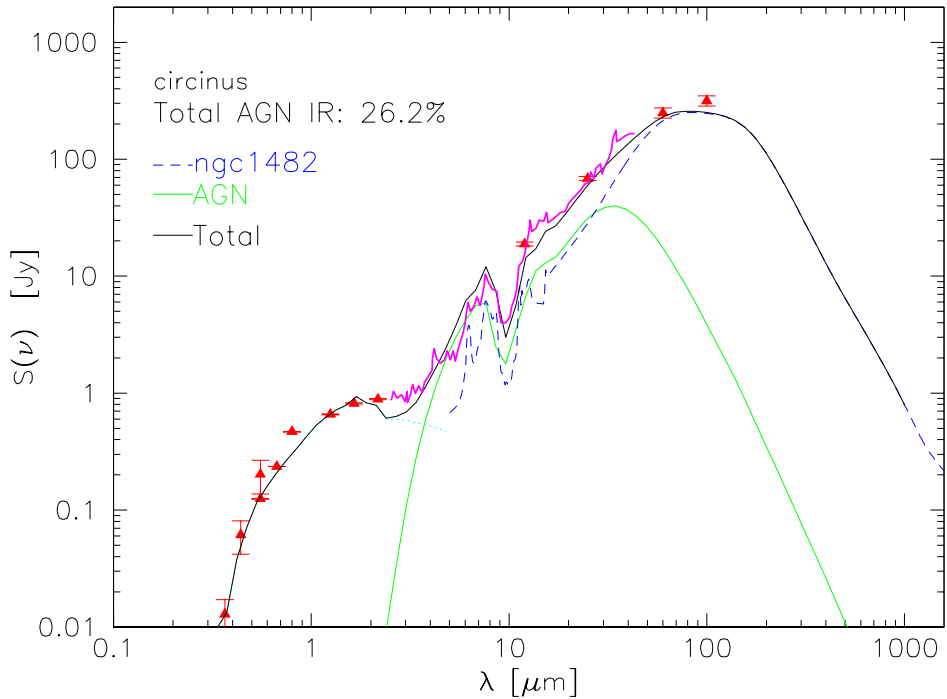}}
\caption{The observed Spectral Energy Distribution of the Circinus galaxy (red triangles) including the $2.5-45$ 
\mums ISO {\sc SWS} spectrum (thick cyan continuous line). The emission has been 
modelled using the NGC  1482 IR template spectrum (dashed line) plus a highly obscured 
AGN torus (thin line). The torus has an optical depth $\tau(9.7)=8.0$, an aperture 
angle of $140^\circ$ and a outer to inner radii ratio of $60$, and an equatorial view. 
The bolometric AGN luminosity turns out to be $\sim 1.78 \times 10^{44}$ erg s$^{-1}$. 
The density profile of dust follows a power-law radial profile with index $\beta=-1$ 
and $\gamma=6.0$. The integrated stellar emission component was taken from the optical 
spectrum of IRAS 19254-7245 (\citealt{berta03}; Fritz et al. in prep.).}
\label{fig:circinus}
\end{figure*}

\begin{figure*}
\centering
\rotatebox{270}{
\includegraphics[height=0.8\textwidth]{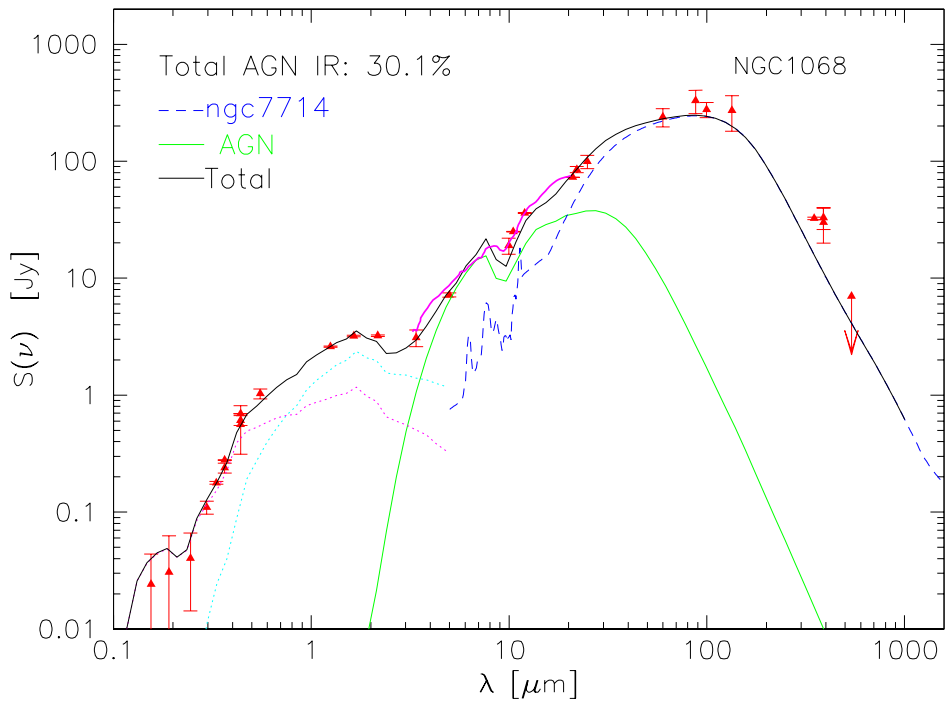}}
\caption{The SED of the Seyfert 2 galaxy NGC  1068, including the {\it ISO} combined {\sc LWS+SWS} spectrum (thick magenta line) taken from Genzel \& Cesarsky (2000). The model includes the contribution of the IR template of the starburst NGC  7714. The assumed stellar spectrum is obtained as a composition of two models of the ULIRG IRAS 20551-4250 (Fritz et al. in preparation). The torus model has an aperture angle of $160^\circ$, an outer-to-inner radius ratio of $20$ and a distribution of the dust density varying both in the radial and in the altitude coordinate ($\beta=-1.0$ and $\gamma=6.0$), with a value of the optical depth at $9.7$ \mums of $8$. The total luminosity is $3.9 \times 10^{45}$ erg s$^{-1}$ and the outer radius $\sim 16.4$ pc.}
\label{fig:NGC 1068_fit}
\end{figure*}

\begin{figure*}
\centering
\rotatebox{270}{
\includegraphics[height=0.8\textwidth]{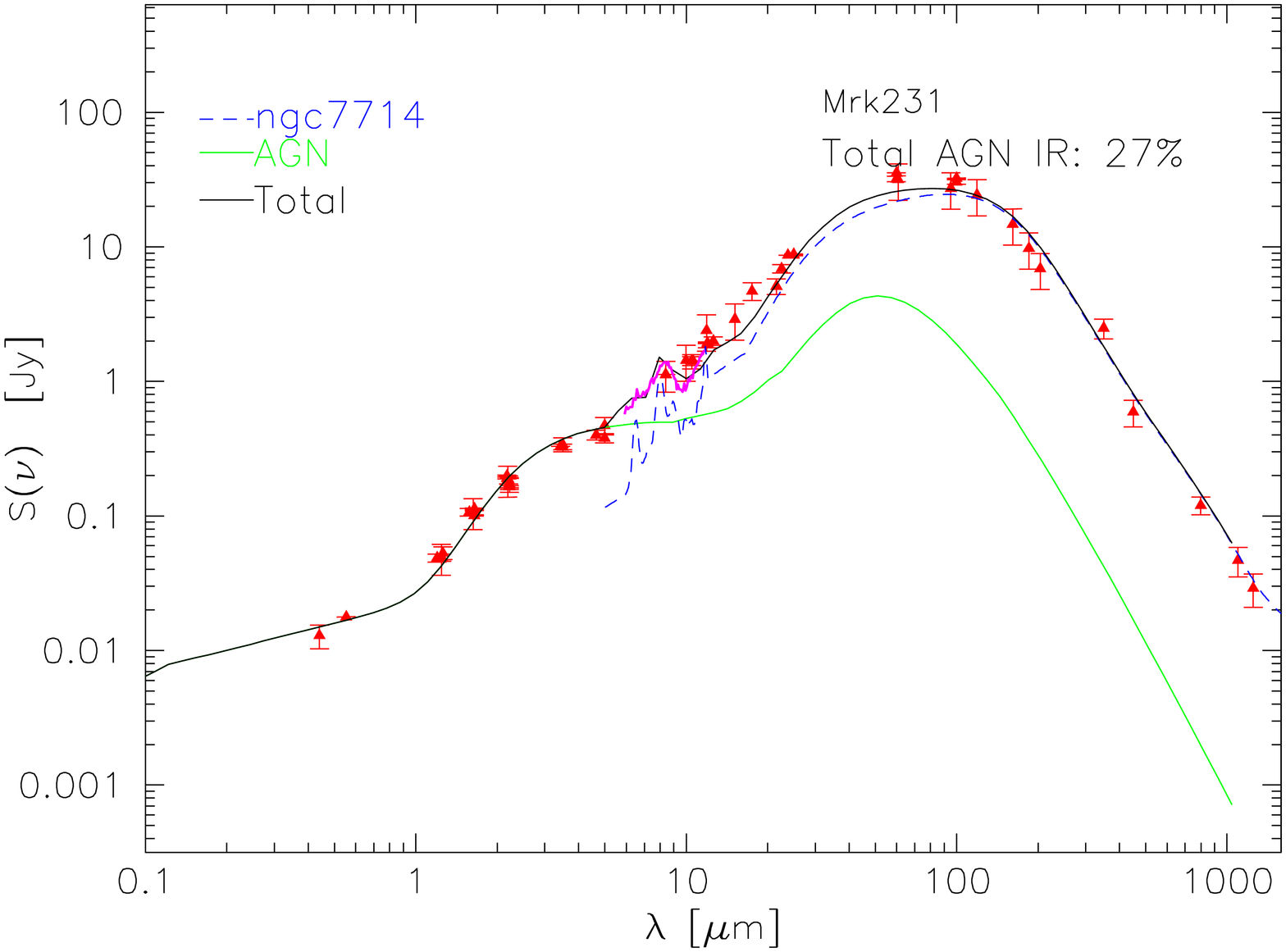}}
\caption{IR SED and spectrum of Mrk  231 versus model fit.
To the far IR part of the SED the spectral template of NGC  5253 has been 
added. This type-1 AGN has a torus with optical depth of $10$, outer radius of $300 \times R_{min}$ with $R_{min}=0.69$ pc and total torus aperture angle of $140^\circ$. The line-of-sight angle with respect to the equatorial plane of the torus is $\Psi=25^\circ$, hence grazing the upper torus edge.}
\label{fig:Mrk 231}
\end{figure*}

\subsection{``Prototypical'' sources}
\label{proto}

In order to demonstrate the accuracy and reliability of our spectral modelling, 
we have studied in some detail the observed SEDs and IR spectra
of three well known nearby objects. 
One of the three objects, Mrk  231, was chosen because it represents a type-1 AGN-dominated ULIRG, while NGC  1068 was chosen as a prototype type-2 AGN-dominated source.
The Circinus galaxy was finally considered as a composite object.  
All three are very well studied, given their proximity and brightness.

Spectroscopic data in the mid-IR from {\it ISO} ({\sc SWS}) were available for the three sources, imposing strong constraints to the torus characteristics. For this reason we built other 
dedicated models with slightly different characteristics with respect to our reference grid 
(see Section \ref{sec:modgrid}), 
to try to fit both the observed photometry and the mid-IR spectrum.

\subsubsection{The Circinus galaxy}

This very nearby ($\sim 4$ Mpc) source displays characteristics typical of a Seyfert 2 object, like for example the very intense forbidden optical lines \citep{oliva94} and a broad component of H$\alpha$ emission \citep{oliva98}. Moreover, indications of current star formation activity close to the central region of this spiral galaxy come from near-IR observations \citep{maiolino98} and from the extended morphology of the H$\alpha$ emission \citep{elmouttie98}.

The photometric datapoints available for this source from the optical to the far-IR and the {\it ISO} {\sc SWS} spectrum ($2.5$ to $45$ \mum; \citealt{moorwood96}, see Fig. \ref{fig:circinus}) strongly constrain the AGN contribution. A large value for the equatorial optical depth 
[$\tau(9.7)=8.0$, corresponding to $A_V\sim 170$ magnitudes] is required in order to depress dust emission between $2$ and $10$ \mums to match the observed spectrum, and a small values for the torus size (internal and external boundaries of $0.2$ and $12$ pc, with an aperture angle of $140^\circ$) is indicated to reproduce the longer wavelength part of the spectrum.  
The estimated mass of torus dust is $\sim 380$ M$_\odot$, a low value ensuing our small best-fit torus size.

Our torus model component has a flared disc geometry and its main features are quite similar to those suggested by \cite{ruiz01}: small torus size (the parameter $h/R_{max}$, where $h$ is the vertical size, is $\sim 0.98$ for this model) and a high visual extinction. The torus sizes of our solution are fairly consistent with their results. The most important difference concerns the relative staburst-AGN IR contribution: \cite{ruiz01} find this ratio to be close to unity, while \cite{moorwood96} estimate the $\sim 90$ per cent of the total IR luminosity to be ascribed to the AGN component. Our best-fit value for the AGN contribution is $\sim 30$ per cent, much closer to that proposed by \cite{rr-craw89}. Differences in the results might be due to aperture
correction effects, since the AGN emission is concentrated in the inner galaxy bulge, while 
the starburst component is more extendend \citep{maiolino98}.

\subsubsection{NGC 1068}

NGC 1068 is a very well studied prototype Seyfert 2 galaxy and the first 
one against which the Unified AGN Model was tested 
(see e.g. \citealt{antonucci84} and \citealt{antonucci85}). 
Among others, studies of the putative dusty torus in this source included the IR polarized imaging by \cite{young96}, the modeling by \cite{efstat1068_95}, and recent mid-IR 
interferometric observations by \citep{jaffe04} spatially resolving 
the dusty torus surrounding 
the active nucleus.

The torus model that better fits the data (see Fig. \ref{fig:NGC 1068_fit}) is one with an outer-to-inner radial ratio of $20$ which,  rescaled to fit the observed datapoints, gives a value of about $17.4$ pc for the external radius for a luminosity of $\sim 3.9\times 10^{45}$ erg s$^{-1}$ of the central source. An optical depth of $\tau(9.7)=8.0$ along the equatorial plane,
together with a non-constant density profile ($\beta$ and $\gamma$ parameters for the densty law are $-1.0$ and $6.0$ respectively), yield a silicate feature which is only slightly in absorption, as seen both in the observed SED and in the {\it ISO} spectrum. The derived optical extinction for such a configuration of the torus is $A_V\sim 170$, corresponding to a hydrogen column density of $6.98\times 10^{23}$ cm$^{-2}$ and a total mass of dust of $1.26 \times 10^3$ M$_\odot$.

The colder component of dust emission is well reproduced by adding the contribution 
of the IR template of the starburst galaxy NGC 1482, contributing some $70$ per 
cent of the total IR luminosity. Our results on the AGN torus characteristics are in 
overall good agreement with previous studies, but indicate a larger
torus than that inferred by the interferometric studies of \cite{jaffe04}, who reveals
the presence of dust at $320$ K in a structure of $3.4$ pc in diameter. Anyway, in the model
adopted to reproduce NGC 1068 emission in the mid-infrared, the dust reaches such a temperature
within a radius of $\sim 8$ pc. 
Considering that the density law in this particular case takes the form:
\begin{equation}
\rho(r,\theta)=\alpha \cdot r^{-1}\cdot e^{-6 \times |\cos(\theta)|}
\end{equation}
we can easily see that the dust density drops in such a way that at $8$ pc
it is $\sim 5$ times lower than that at $\sim 1.7$ pc (radius
reported by \citealt{jaffe04}). Our findings are, therefore, in marginal agreement
with the obsevations.

\subsubsection{Mrk  231}

This object is one of the most luminous ULIRGs in the local universe \citep{soifer84}, and also one of the most studied at all wavelength. Both {\it ROSAT} and {\it ASCA} observations found clear evidence of a starburst activity combined to that of the AGN (\citealt{imanishi99}; \citealt{turner99}). The analysis based on {\it Beppo-SAX} and {\it XMM-Newton} performed by \cite{braito04} indicates the presence of a very thick screen
with an hydrogen column density of $\sim 2 \times 10^{24}$ cm$^{-2}$ covering the primary X-ray emission.

The torus model is very strictly constrained by the observational 
data in the $1$ to $5$ \mums interval.
To reproduce the observed optical-to-IR SED (shown in Fig. \ref{fig:Mrk 231}) we need a direct view of the central power source. The observed SED requires an additional contribution by colder dust at $\lambda>10$ \mum, which we reproduce using the NGC 5253 IR SED. 

The torus model has $\tau(9.7)\simeq 10.0$ (corresponding to $A_V\simeq 210$)
and a mass of dust confined in the molecular torus of about $2.7\times 10^6$ M$_\odot$, 
$R_{max}/R_{min}=300$ and the torus aperture angle $\Theta=140^\circ$. Such a high 
value for the optical depth implies a hydrogen column density of $\sim 9\times 10^{23}$ 
cm$^{-2}$, along the equatorial plane of the torus. The model AGN bolometric
luminosity is $2.8\times 10^{45}$ erg s$^{-1}$, corresponding to an outer radius 
of $\sim 200$ pc.

Our fit (whose parameters are summarized in Table 2) implies an AGN contribution to the IR flux of about $27$ per cent, two thirds 
of the emission being attributed to the host galaxy.
This is consistent with the results by \cite{braito04}, that found the
starburst luminosity to be significantly higher than that of the AGN
from a comparison of the X-ray and the bolometric flux and assuming a standard 
type-1 AGN emission.

\subsection{Further Applications}
\label{scale}

In this Section we study the observed broad band SEDs of a sample of 
40 type-1 Seyfert galaxies and quasars and 15 type-2 Seyferts,
following the same procedure as in the previous Section but making
use of our standard grid of models (see Section \ref{sec:modgrid}). 
The aim is to obtain information about physical properties of the
circum-nuclear dust distributions in a sample of local AGNs,
as well as on the relative contribution of the host galaxy.

\begin{figure*}
\centering
\includegraphics[height=1.30\textwidth]{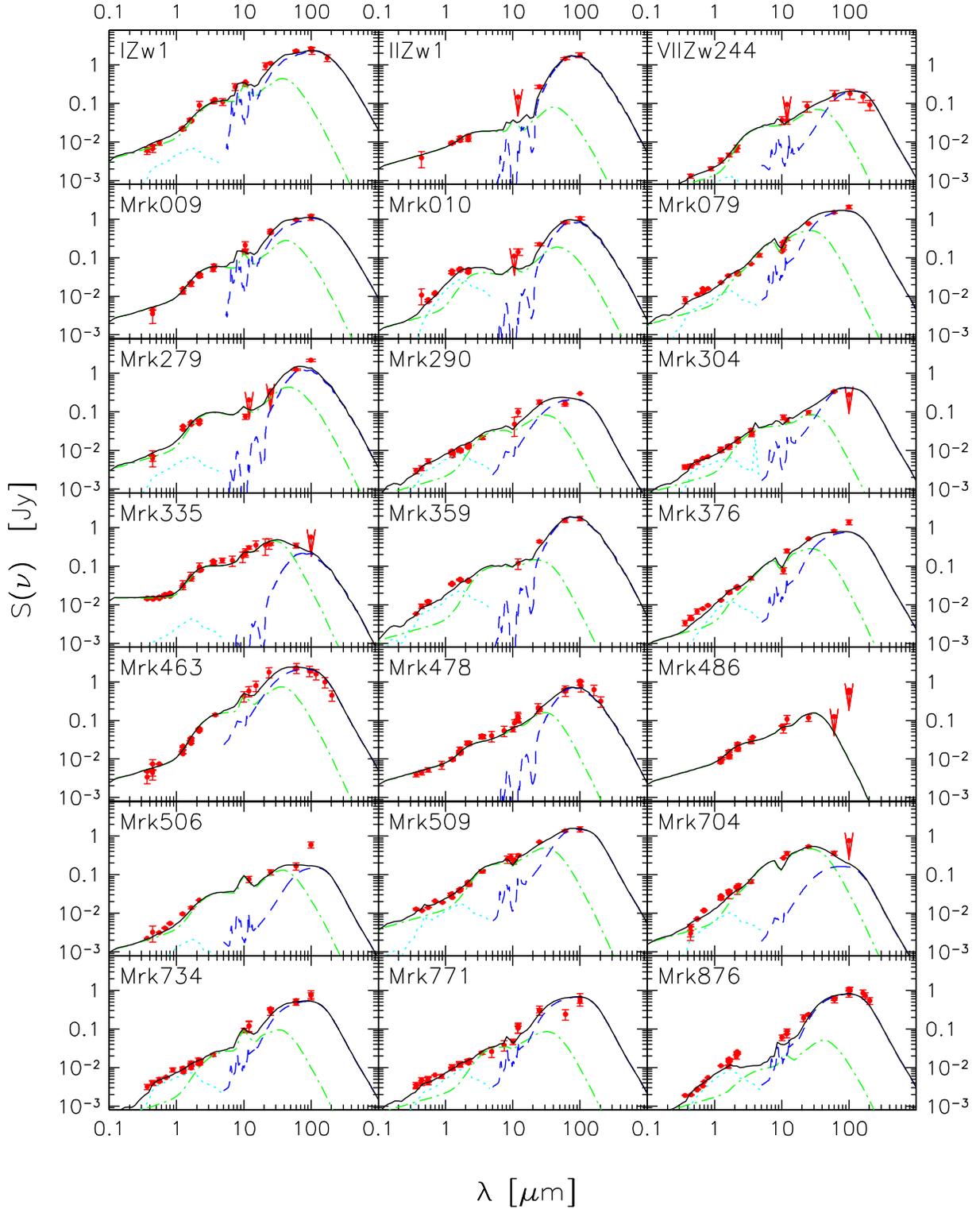}
\caption{Comparison of our best-fit models with observational data for the type-1 AGN.
Red triangles are the observed data points; the solid black line is the total model SED;
green line is the AGN emission; the blue dashed line denotes the starburst emission; stellar emission by the host galaxy is shown as dashed lines.}
\label{fig:QSO1_1}
\end{figure*}

\begin{figure*}
\centering
\includegraphics[height=1.30\textwidth]{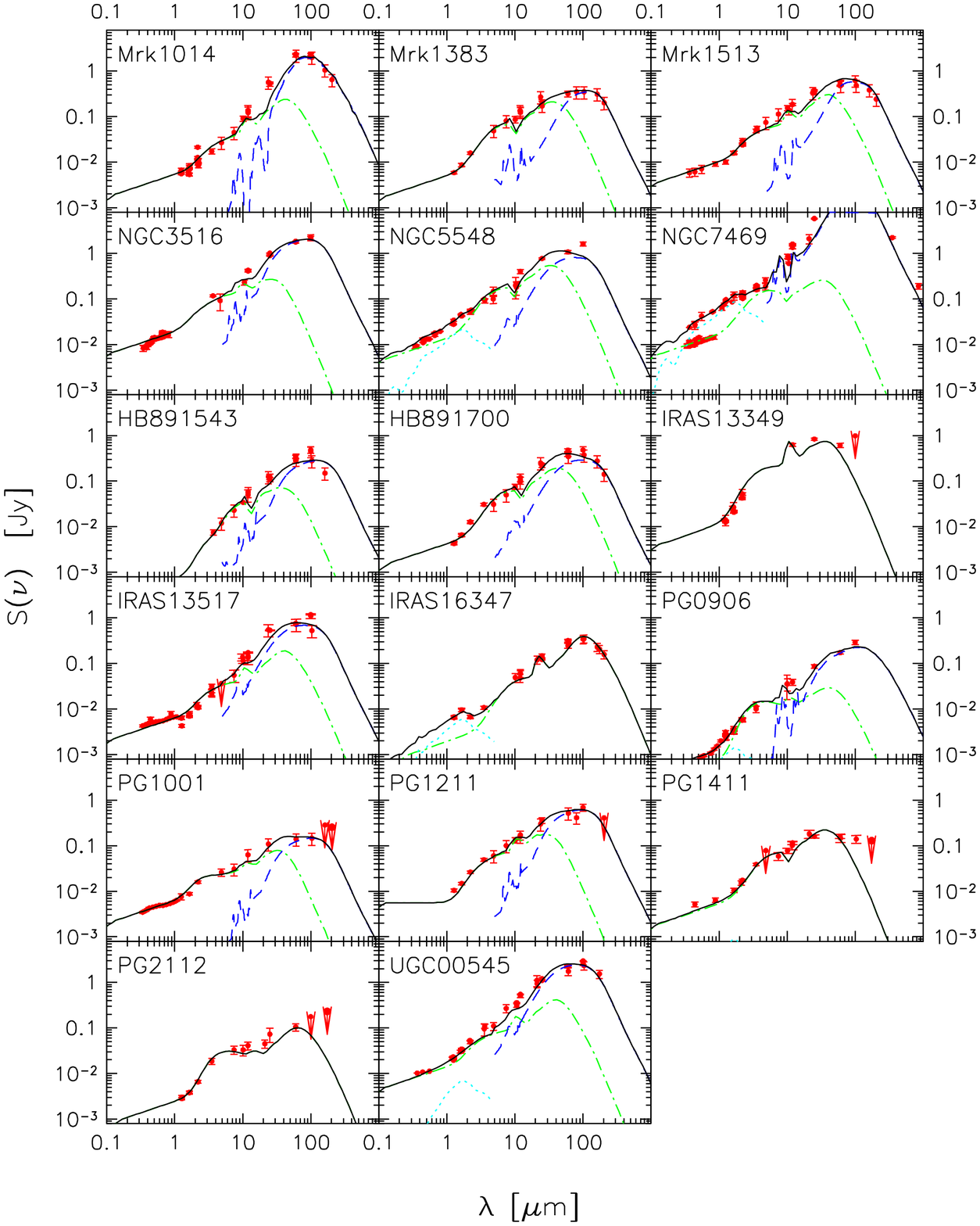}
\caption{Comparison of our best-fit models with observational data for the type-1 AGN. Meaning
of symbols and line is the same as for Fig. \ref{fig:QSO1_1}.}
\label{fig:QSO1_2}
\end{figure*}

In this exercise, we have taken particular care in selecting
photometic measurements within homogeneous apertures, as much as possible. 
An example is given by our fit of the type-1 AGN NGC  7469 (Fig. \ref{fig:QSO1_2}), 
for which photometric data with an aperture including only the very central 
regions are used to constrain the intensity of the AGN component, 
while the total intensity fluxes include also the host galaxy contribution.

\subsubsection{Type-1 AGN}

A total of $40$ type-1 AGN and quasars, with a good coverage of 
photometric datapoints, have been fitted with a combination of stellar, 
IR starburst and AGN emissions. The objects were chosen from various AGN 
samples with the only requirement to have a sufficient number of observational datapoints.
Our first attempt was to reproduce the whole
optical-IR observed SEDs with a pure AGN emission, trying all models in the grid
and assessing their merit with the $\chi^2$ test. 
For those objects with unacceptably large $\chi^2$ and obvious far-IR excess
we added starburst and normal galaxy components in the far-IR and optical, respectively.

In all cases the combination of the three emission components reproduced well all 
the observed SEDs. Figs. \ref{fig:QSO1_1} and \ref{fig:QSO1_2} show the observed 
(red triangles) and model (black solid line) SEDs. The figures also report
the various components of the model: the IR starburst and AGN contributions 
on blue dashed and green solid lines, respectively, while
the stellar emission is shown as dotted lines.

\begin{table*}
\small
\begin{tabular}{l c c c c c c c c c c c c c}
\hline
\hline
 Obj       &	z     &    L$^{AGN}$	& R$_{min}$ & R$_{Max}$& $\Theta$ & $\beta$ & $\gamma$ & $\tau(9.7)$ & A$_V$  &   N$_H$   & M$_{dust}$ & AGN & $\Psi$ \\
           &          & [$10^{46}$erg s$^{-1}$]&   [pc]   &   [pc]   & $^\circ$ &	   &	      & 	    & [mag]  & $cm^{-2}$ &  M$_\odot$ & [\%] & $^\circ$ \\
\hline
3C249.1        &0.31120& $0.89$ & $1.23$ &$36.8$ &100&-1.0&6.0& $10.0$ & $214.5$ & $8.68\times 10^{23}$ & $6.85\times 10^3$ &72.2&  0\\
IRAS 13349     &0.10764& $0.79$ & $1.16$ & $116$ &140& 0  & 0 &  $1.0$ &  $21.6$ & $8.79\times 10^{22}$ & $1.13\times 10^5$ &100 &  0\\
IRAS 13517     &0.08820& $0.32$ & $1.29$ & $129$ &100&-1.0&6.0& $10.0$ & $214.5$ & $8.68\times 10^{23}$ & $6.20\times 10^4$ &29.6& 20\\
IRAS 16347     &1.33400& $7.94$ & $3.66$ &$1100$ &140& 0  & 0 &  $2.0$ &  $42.7$ & $1.73\times 10^{23}$ & $1.99\times 10^7$ &100 & 20\\
$[$HB89$]$1543+489 &0.40000& $0.56$ & $0.98$ & $29.2$ &140&-1 & 0 &$10.0$ & $214.5$ & $8.68\times 10^{23}$ & $3.12\times 10^4$ &55.3& 0\\
$[$HB89$]$1700+518 &0.29200& $0.79$ & $1.69$ & $50.7$ &140&-1 & 0 & $6.0$ & $129.1$ & $5.23\times 10^{23}$ & $5.42\times 10^4$ &68.1& 0\\
IZw1	       &0.06114& $0.25$ & $0.65$ &$65.2$ &100& 0  & 0 &  $2.0$ &  $43.5$ & $1.76\times 10^{23}$ & $5.80\times 10^4$ &35.0&  0\\
IIZw1	       &0.05434& $0.14$ & $0.49$ & $147$ &100&-0.5&6.0&  $1.0$ &  $21.9$ & $8.94\times 10^{22}$ & $1.47\times 10^3$ &14.3&  0\\
VIIZw244       &0.13100& $0.14$ & $0.49$ &$14.7$ &140& 0  & 0 &  $6.0$ & $129.0$ & $5.23\times 10^{23}$ & $1.16\times 10^4$ &76.2&  0\\
Mrk 009	       &0.03987& $0.07$ & $0.35$ & $104$ &100&-0.5& 0 &  $1.0$ &  $21.9$ & $8.94\times 10^{22}$ & $4.16\times 10^4$ &47.6&  0\\
Mrk 010	       &0.02925& $0.03$ & $0.21$ &$61.8$ &100&-0.5& 0 &  $3.0$ &  $64.9$ & $2.64\times 10^{23}$ & $4.36\times 10^4$ &53.3&  0\\
Mrk 079	       &0.02219& $0.01$ & $0.15$ & $4.5$ &140&-1  & 0 & $10.0$ & $214.5$ & $8.68\times 10^{23}$ & $7.30\times 10^2$ &46.5&  0\\
Mrk 231	       &0.04217& $0.28$ & $0.69$ & $207$ &140&-0.5& 0 &  $6.0$ & $128.8$ & $5.22\times 10^{23}$ & $1.34\times 10^6$ &27.5&  0\\
Mrk 279	       &0.03045& $0.03$ & $0.23$ &$69.4$ &100&-0.5& 0 &  $3.0$ &  $64.9$ & $2.64\times 10^{23}$ & $5.48\times 10^4$ &49.1&  0\\
Mrk 290	       &0.02958& $0.01$ & $0.15$ & $4.4$ &140&-0.5& 0 &  $6.0$ &$129.43$ & $5.25\times 10^{23}$ & $7.01\times 10^2$ &53.8&  0\\
Mrk 304	       &0.06576& $0.15$ & $0.51$ &$15.2$ &100&-1.0&6.0& $10.0$ & $214.5$ & $8.67\times 10^{23}$ & $1.16\times 10^3$ &64.5& 20\\
Mrk 335	       &0.02578& $0.18$ & $0.55$ &$16.4$ &100&-0.5&6.0&  $6.0$ & $129.5$ & $5.26\times 10^{23}$ & $1.38\times 10^3$ &88.3&  0\\
Mrk 359	       &0.01738& $0.01$ & $0.10$ &$31.0$ &140& 0  & 0 &  $6.0$ & $127.9$ & $5.16\times 10^{23}$ & $4.69\times 10^4$ &59.8&  0\\
Mrk 376	       &0.05598& $0.06$ & $0.33$ & $9.8$ &140&-1  & 0 & $10.0$ & $214.5$ & $8.68\times 10^{23}$ & $3.49\times 10^3$ &57.6&  0\\
Mrk 463         &0.05035& $0.14$ & $0.49$ & $148$ &140& 0  & 0 &  $6.0$ & $127.9$ & $5.16\times 10^{23}$ & $1.08\times 10^6$ &71.0&  0\\
Mrk 478	       &0.07905& $0.32$ & $0.73$ &$21.9$ &100&-1  &6.0& $10.0$ & $214.5$ & $8.68\times 10^{23}$ & $2.43\times 10^3$ &50.8&  0\\
Mrk 486	       &0.03893& $0.08$ & $0.37$ &$11.0$ &140&-0.5&6.0& $10.0$ & $214.3$ & $8.67\times 10^{23}$ & $1.03\times 10^3$ &100 &  0\\
Mrk 506	       &0.04303& $0.04$ & $0.26$ &$25.9$ &100&0.0 & 0 &  $3.0$ &  $64.5$ & $2.62\times 10^{23}$ & $1.18\times 10^4$ &86.4&  0\\
Mrk 509	       &0.03440& $0.09$ & $0.40$ &$11.9$ &140&-0.5& 0 &  $6.0$ & $129.4$ & $5.25\times 10^{23}$ & $5.20\times 10^3$ &65.5&  0\\
Mrk 704	       &0.02923& $0.04$ & $0.25$ & $7.4$ &140&-1  & 0 & $10.0$ & $214.5$ & $8.68\times 10^{23}$ & $1.97\times 10^3$ &90.0&  0\\
Mrk 734	       &0.02923& $0.03$ & $0.21$ &$61.8$ &140& 0  & 0 &  $2.0$ &  $42.7$ & $1.73\times 10^{23}$ & $6.29\times 10^4$ &51.5&  0\\
Mrk 771	       &0.06301& $0.02$ & $0.20$ & $5.8$ &140&-0.5& 0 & $10.0$ & $214.3$ & $8.67\times 10^{23}$ & $6.51\times 10^3$ &41.9&  0\\
Mrk 876	       &0.12900& $0.28$ & $1.22$ &$122$  &100&-1  &6.0& $10.0$ & $213.7$ & $8.64\times 10^{23}$ & $5.52\times 10^4$ &23.5&  0\\
Mrk 1014        &0.16301& $0.89$ & $1.23$ & $123$ &140&-1  &6.0& $10.0$ & $213.6$ & $8.63\times 10^{23}$ & $6.51\times 10^4$ &33.2&  0\\
Mrk 1383        &0.08647& $0.26$ & $0.67$ &$20.0$ &140&-0.5& 0 & $10.0$ & $214.3$ & $8.67\times 10^{23}$ & $2.42\times 10^4$ &79.7&  0\\
Mrk 1513        &0.06298& $0.28$ & $1.21$ &$121$  &100&-1  &6.0& $10.0$ & $213.6$ & $8.63\times 10^{23}$ & $5.40\times 10^4$ &70.7&  0\\
NGC 3516        &0.00884& $0.01$ & $0.12$ & $3.7$ &100&-1  &6.0& $10.0$ & $214.5$ & $8.68\times 10^{23}$ & $7.0 \times 10^1$ &32.8&  0\\
NGC 5548        &0.01717& $0.03$ & $0.23$ & $6.6$ &140&-0.5& 0 & $10.0$ & $214.3$ & $8.67\times 10^{23}$ & $2.60\times 10^3$ &67.5&  0\\
NGC 7469        &0.01632& $0.03$ & $0.23$ & $6.9$ &100&-1  & 0 & $10.0$ & $214.5$ & $8.68\times 10^{23}$ & $1.25\times 10^3$ & 8.8&  0\\
PG0804+761     &0.09990& $0.85$ & $1.20$ &$36.0$ &100&-1.0&6.0& $10.0$ & $214.5$ & $8.68\times 10^{23}$ & $6.54\times 10^3$ &92.1&  0\\
PG0906+484     &0.11800& $0.16$ & $0.52$ &$15.5$ &100&0.0 & 0 & $10.0$ & $214.0$ & $8.65\times 10^{23}$ & $6.27\times 10^3$ &28.7&  0\\
PG1001+054     &0.16100& $0.96$ & $1.27$ &$38.1$ &100&-0.5&6.0&  $6.0$ & $214.3$ & $8.67\times 10^{23}$ & $4.45\times 10^3$ &72.5&  0\\
PG1211+143     &0.08090& $0.59$ & $0.77$ &$23.0$ &100&-1  &6.0& $10.0$ & $214.5$ & $8.68\times 10^{23}$ & $4.53\times 10^3$ &63.8&  0\\
PG1411+442     &0.08960& $0.28$ & $0.69$ &$20.7$ &140&-0.5& 0 & $10.0$ & $214.3$ & $8.67\times 10^{23}$ & $2.59\times 10^4$ &100 &  0\\
PG2112+059     &0.46600& $2.51$ & $2.06$ & $206$ &100&-0.5& 0 & $10.0$ & $214.2$ & $8.67\times 10^{23}$ & $1.66\times 10^6$ &100 &  0\\
UGC00545       &0.06114& $0.43$ & $1.50$ &$150$  &100&-1  &6.0& $10.0$ & $213.7$ & $8.64\times 10^{23}$ & $8.36\times 10^4$ &29.5&  0\\
\hline
\end{tabular}
\normalsize
\caption{Best-fit physical parameters of the torus model for all type-1 AGN. In the order, object name, redshift, AGN luminosity, inner and outer boundary radius, the full-opening angle of the torus, the radial and angular density dependency, the optical depth at $9.7$ \mum, the extinction A$_V$, the hydrogen column density N$_H$, the mass of dust and the percentage contribution of the AGN. The latter is computed in the range $5\div 1000$ \mum. The viewing angle $\Psi$, measured with respect to the $z$ axis, is in general taken to be $0^\circ$, since the models show in general very little dependence on this parameter for type-1 AGN.}
\label{tab:sy1_res}
\end{table*}

Table \ref{tab:sy1_res} summarizes the physical parameters derived for the 41 type-1 AGN under investigation (including Mrk 231). In the order, the columns report the object name, redshift $z$, the luminosity of the central power source, the inner and outer torus radii, the full-opening angle of the torus, the radial and angular dust density dependences (parameters $\beta$ and $\gamma$, respectively), the optical depth at $9.7$ \mum, the extinction A$_V$, the hydrogen column density N$_H$, the mass of dust and the percentage contribution of the AGN to the IR bolometric emission. Note that $\tau(9.7)$, $A_V$ and $N_H$ are measured between the inner and the outer radius in the torus equatorial plane.

For the majority of the objects ($27$ out of $40$), the contribution of the AGN torus to the total IR emission is larger than $50$ per cent. The SEDs of five objects in particular, Mrk 486, IRAS13349, IRAS16347, PG1411+442 and PG2112+059 can be entirely reproduced using a pure AGN model, fitting of their observed SED does not require significant cold dust emission. 

As for the short-wavelength optical-UV part of the spectrum,
there is a remarkable complementarity of the spectral shapes for the AGN and host galaxy:
the local minima of one component correspond to the maxima of the other, including the $9.7$ \mums emission/absorption. This coincidental fact allows us to assess with some precision
the contribution by the host galaxies to optical spectrum.
Quite often in these moderate luminosity Seyfert-1 galaxies the latter appears to be
important (20 cases out of 40).

\begin{table*}
\small
\begin{tabular}{l c c c c c c c c c c c c c}
\hline
\hline
 Obj       &	z     &   L$^{AGN}$    & R$_{min}$ & R$_{Max}$& $\Theta$ & $\beta$ & $\gamma$ & $\tau(9.7)$ & A$_V$  &	N$_H$	& M$_{dust}$ & AGN &$\Psi$ \\
           &          &[$10^{46}$erg s$^{-1}$]&   [pc]   &   [pc]   & $^\circ$ &	  &	     &  	   & [mag]  & $cm^{-2}$ &  M$_\odot$ & [\%] & $^\circ$ \\
\hline
05189-2524 & 0.04256 & $2.50$ & $2.06$ & $61.8$ &100&-0.5&6.0 &  $6.0$ & $129.2$ & $5.24\times10^{23}$ & $1.95\times 10^5$ & 50.3& 80\\
BGC5506    & 0.00618 & $0.03$ & $0.23$ & $ 6.9$ &140& 0  & 0  &  $3.0$ & $65.1 $ & $2.65\times10^{23}$ & $1.19\times 10^3$ & 37.2& 90\\
Circinus   & 0.00145 & $0.02$ & $0.20$ & $12.0$ &140&-1  &6.0 &  $8.0$ & $171.8$ & $6.96\times10^{23}$ & $3.81\times 10^2$ & 26.2& 90\\ 
Mrk 003	   & 0.01351 & $0.07$ & $0.34$ & $10.3$ &140& 0  & 0  &  $6.0$ & $129.0$ & $5.23\times10^{23}$ & $5.47\times 10^3$ & 70.5& 30\\
Mrk 078	   & 0.03715 & $0.08$ & $0.37$ & $11.0$ &140&-0.5& 0  &  $6.0$ & $129.4$ & $5.25\times10^{23}$ & $4.46\times 10^3$ & 32.7& 30\\
Mrk 1066    & 0.01202 & $0.04$ & $0.26$ & $ 7.8$ &140&-0.5& 0  &  $6.0$ & $129.4$ & $5.25\times10^{23}$ & $2.23\times 10^2$ & 25.5& 40\\
Mrk 273	   & 0.03778 & $0.40$ & $0.82$ & $24.6$ &140&-1  &6.0 & $10.0$ & $214.5$ & $8.68\times10^{23}$ & $2.84\times 10^2$ & 5.8 & 90\\
NGC 0262    & 0.01503 & $0.12$ & $0.45$ & $13.5$ &100&-1  &6.0 &  $6.0$ & $129.6$ & $5.26\times10^{23}$ & $7.20\times 10^1$ & 50.2& 90\\
NGC 1068    & 0.00379 & $0.40$ & $0.82$ & $16.4$ &160&-1  &6.0 &  $8.0$ & $172.1$ & $6.98\times10^{23}$ & $1.26\times 10^3$ & 30.1& 70\\
NGC 1365    & 0.00546 & $0.02$ & $0.17$ & $52.0$ &140& 0  & 0  &  $3.0$ & $64.0$  & $2.59\times10^{23}$ & $7.50\times 10^4$ & 43.3& 30\\
NGC 1386    & 0.00289 & $0.01$ & $0.09$ & $ 2.8$ &140& 0  & 0  &  $3.0$ & $65.1$  & $2.65\times10^{23}$ & $3.96\times 10^2$ & 39.2& 90\\
NGC 1614    & 0.01594 & $1.00$ & $1.30$ & $39.0$ &100&-1  &6.0 & $10.0$ & $214.5$ & $8.68\times10^{23}$ & $7.69\times 10^3$ & 28.6& 90\\
NGC 2110    & 0.00779 & $0.02$ & $0.16$ & $49.1$ &100& 0  & 0  &  $3.0$ & $64.0 $ & $2.59\times10^{23}$ & $5.36\times 10^4$ & 32.9& 90\\
NGC 4507    & 0.01180 & $0.01$ & $0.15$ & $46.4$ &140& 0  & 0  &  $3.0$ & $64.0 $ & $2.59\times10^{23}$ & $3.75\times 10^4$ & 45.9& 90\\
NGC 5506    & 0.00618 & $0.04$ & $0.30$ & $29.9$ &100&-1  &6.0 &  $6.0$ & $129.2$ & $5.25\times10^{23}$ & $2.03\times 10^3$ & 36.9& 80\\
NGC 6240    & 0.02448 & $0.63$ & $1.03$ & $31.0$ &100&-1  &6.0 & $10.0$ & $214.5$ & $8.68\times10^{23}$ & $4.84\times 10^3$ & 14.8& 90\\
NGC 7582    & 0.00525 & $0.04$ & $0.26$ & $ 7.8$ &100&-1  &6.0 &  $6.0$ & $129.6 $& $5.26\times10^{23}$ & $1.90\times 10^2$ & 50.3& 90\\
\hline														        
\end{tabular}													        
\normalsize													        
\caption{Summary of the main features of the torus model used to reproduce the SED of the $17$ type-2 objects
(including Circinus and NGC 1068). $\beta$ and $\gamma$ are the parameters entering the density law, $\Theta$ is the full-opening angle of the torus, $\tau(9.7)$, N$_H$ and $A_V$ refer to the equatorial plane of the torus.}
\label{tab:sy2_res}
\end{table*}

\begin{figure*}
\centering
\includegraphics[height=1.3\textwidth]{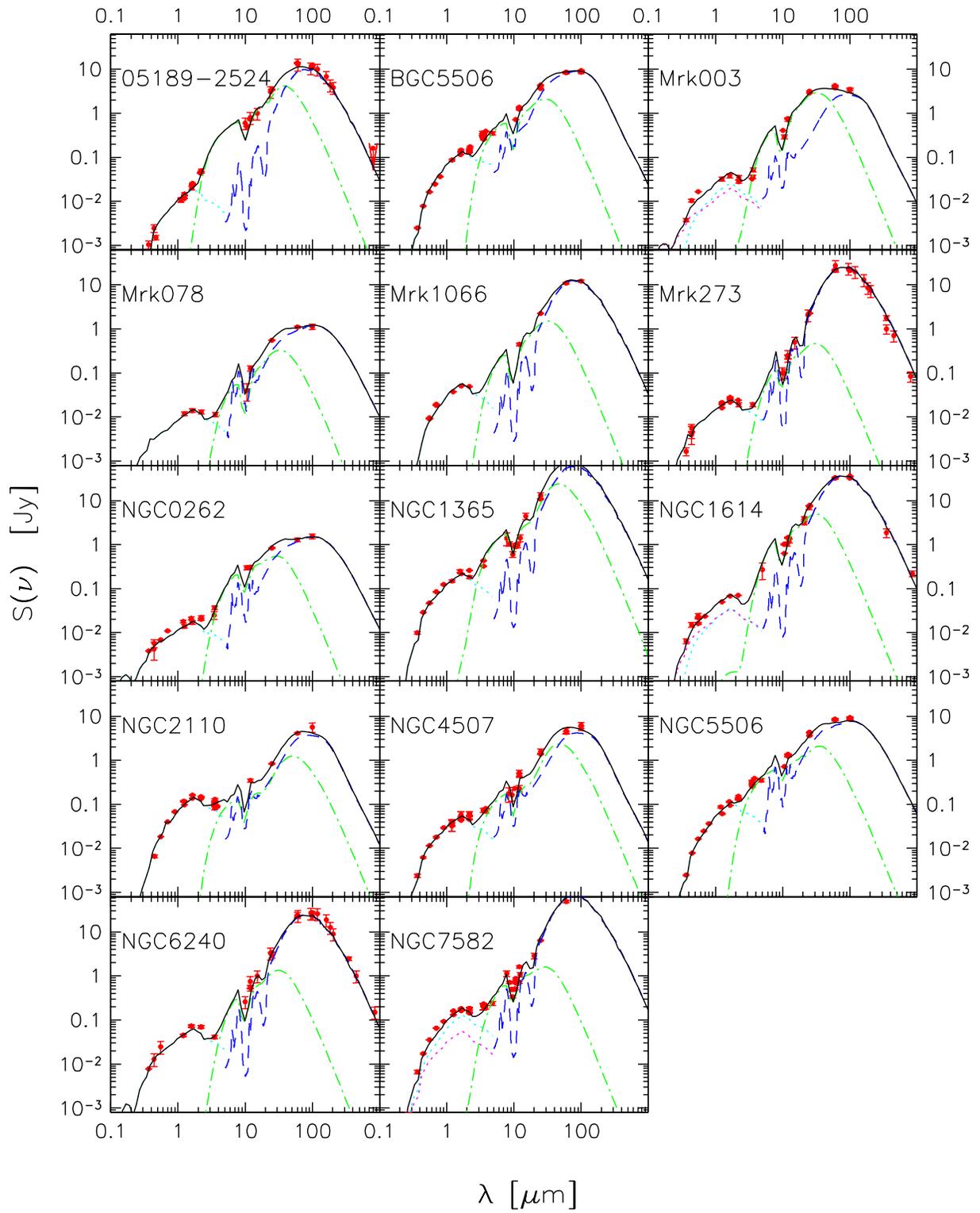}
\caption{Spectral models for the type-2 AGN. Colour and line coding is the same as in the previous figures.}
\label{fig:sey2}
\end{figure*}

Our model solutions for the type-1 objects in this sample tends to favour 
dust tori in which the density has a radial gradient: 32 of the 40 
sample sources require a power-law density profile decreasing with radius 
(i.e. $\beta=-0.5$ or $\beta=-1.0$), and 6 of them also favour an 
angular dependence ($\gamma=6.0$). 

Typical torus sizes are $R_{max}<100\ pc$.
Only for $9$ objects an outer radius barely above $100$ pc is indicated, with the exception of the highest redshift one, IRAS 16347+7037,  
that has a very luminous central source ($\sim 8\times 10^{46}$ erg s$^{-1}$) 
and whose external radius is found to be around $1$ kpc.
As for the torus geometry, it is remarkable that the majority of our sample
sources are well fitted by an high total aperture angle ($\Theta=140^\circ$),
implying an high dust covering factor. 

For the large majority of cases, the values of the mass of dust range from few tens to few tens of thousands solar masses, orders of magnitudes lower than the typical masses of supermassive black hole. Striking exception is again the mass of dust computed for IRAS 16347+7037 found to reach $\sim 10^7$ M$_{\odot}$. 

Most important, the best-fit values of the optical depth at $9.7$ \mums were found 
to be larger than $1$ for almost all the objects, corresponding to
visual extinction values larger than $22$ magnitudes and hydrogen 
column densities in the torus that must exceed $\sim 10^{23}$ cm$^{-2}$.

\subsubsection{Type-2 AGN}

The same fitting procedure has been used to analyse the observed 
broad band SEDs of a small sample of 15 type-2 AGN.
Table \ref{tab:sy2_res} summarizes the main best-fit parameters
(including those for Circinus and NGC 1068). The results of our SED modelling 
(shown in Fig. \ref{fig:sey2}) differ in some aspects 
from those for type-1 objects. 

Again, all observed SEDs can be very well reproduced by combining stellar, starburst and AGN emission. Note that, in this case, the optical and near-IR emission is entirely due to the host galaxy's stellar contribution, since the AGN emission is totally suppressed at wavelengths shorter than $2-3$ \mum. 
None of the type-2 objects can be reproduced with torus emission alone: they all require a substantial starburst component. An Arp 220 starburst template better reproduced the far-IR data in eight out of $15$ cases. 

\begin{figure*}
\centering
\begin{tabular}{l l}
\includegraphics[height=.450\textwidth]{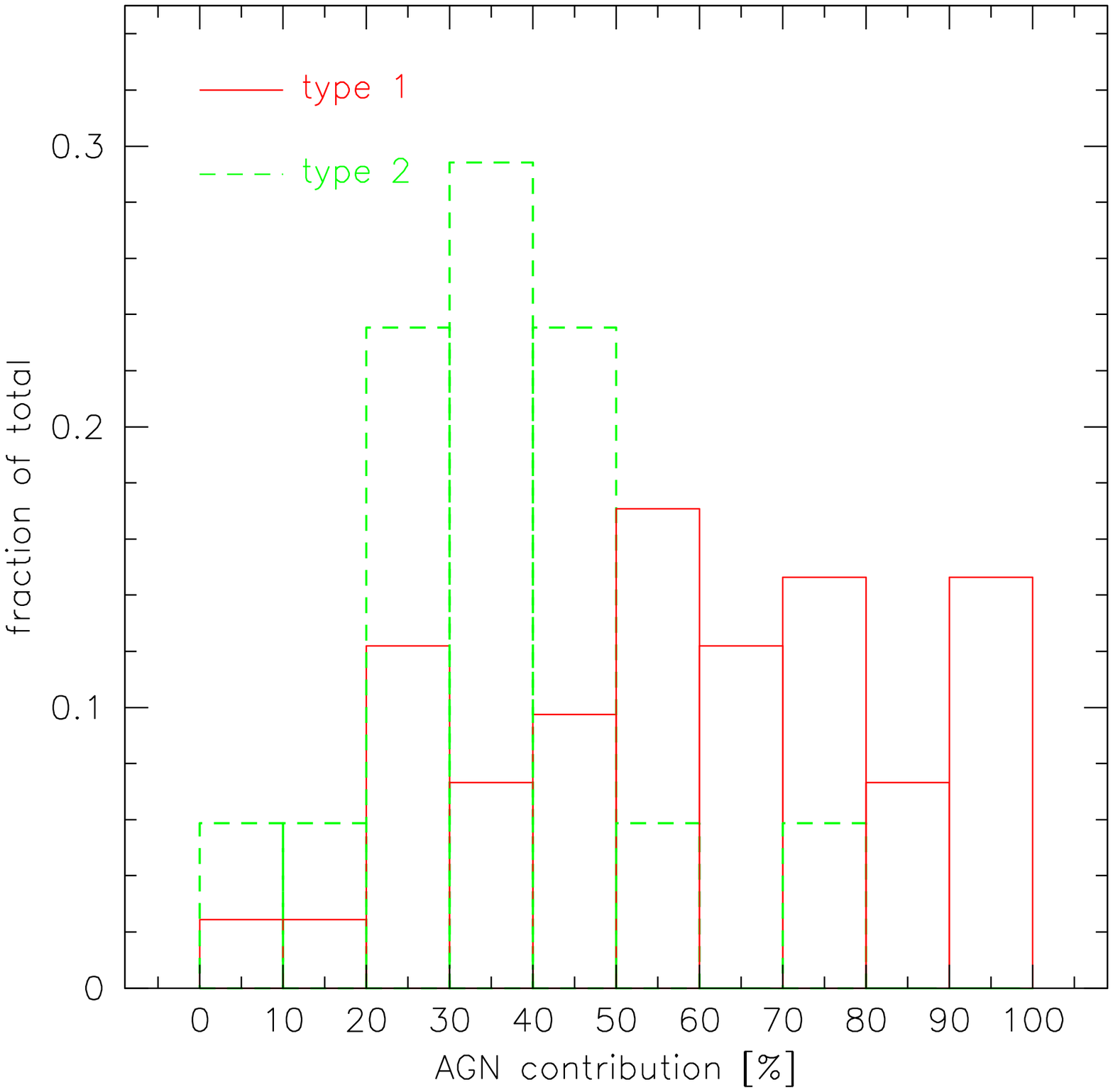} &
\includegraphics[height=.450\textwidth]{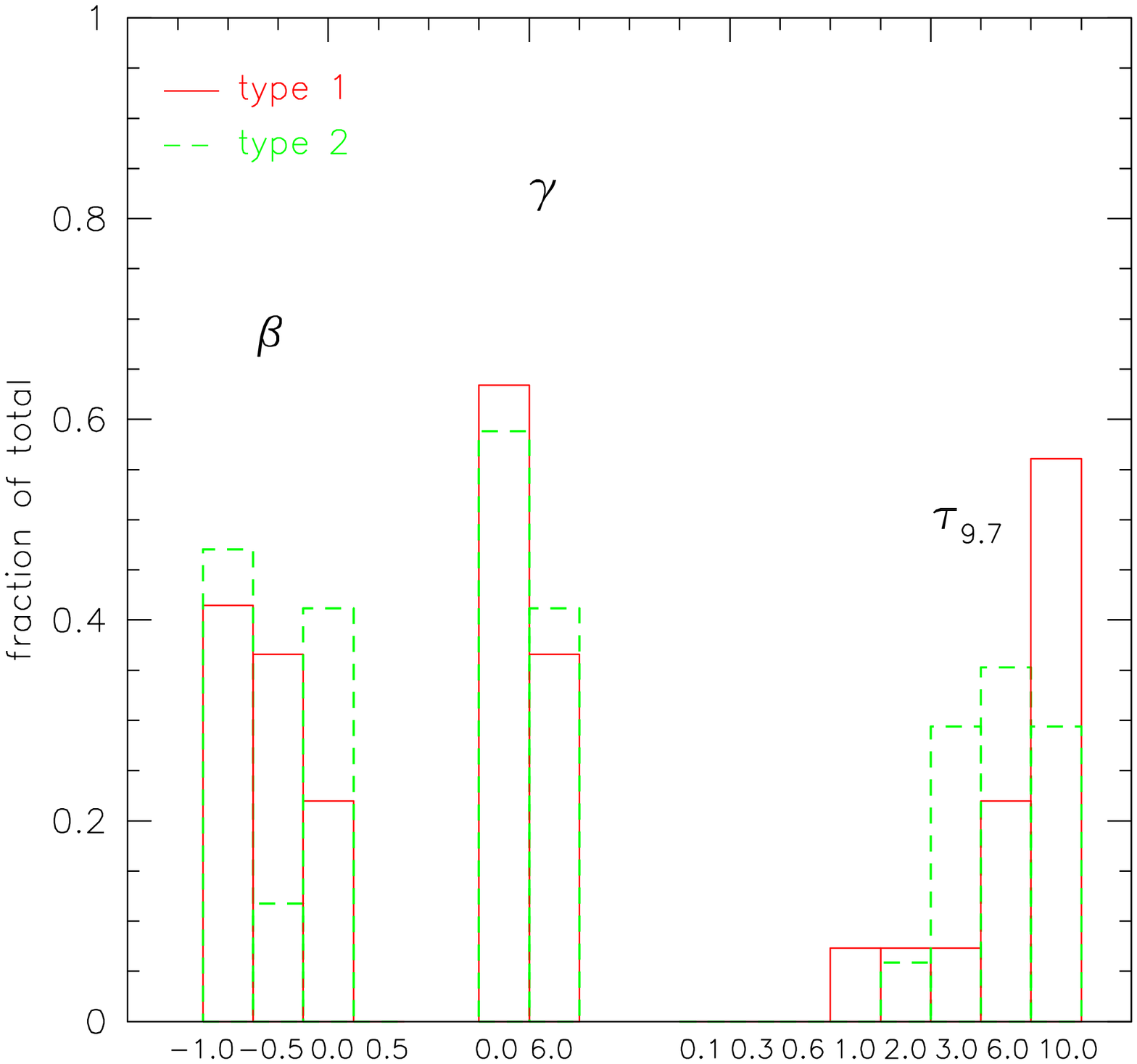} \\
\end{tabular}
\caption{Left panel: fractional contribution of the AGN component to the IR bolometric ($5-1000$\mum) luminosity versus the redshift, for type-1 AGN (red 4-leg stars) and type-2 (green triangles) AGN.
Right panel: summary of the main characteristics of the torus best-fit parametric solutions, with a comparison of type-1 (red continuous line) and type-2 (green dashed line) sources.}
\label{fig:props}
\end{figure*}

\begin{figure*}
\centering
\begin{tabular}{l l}
\includegraphics[height=.450\textwidth]{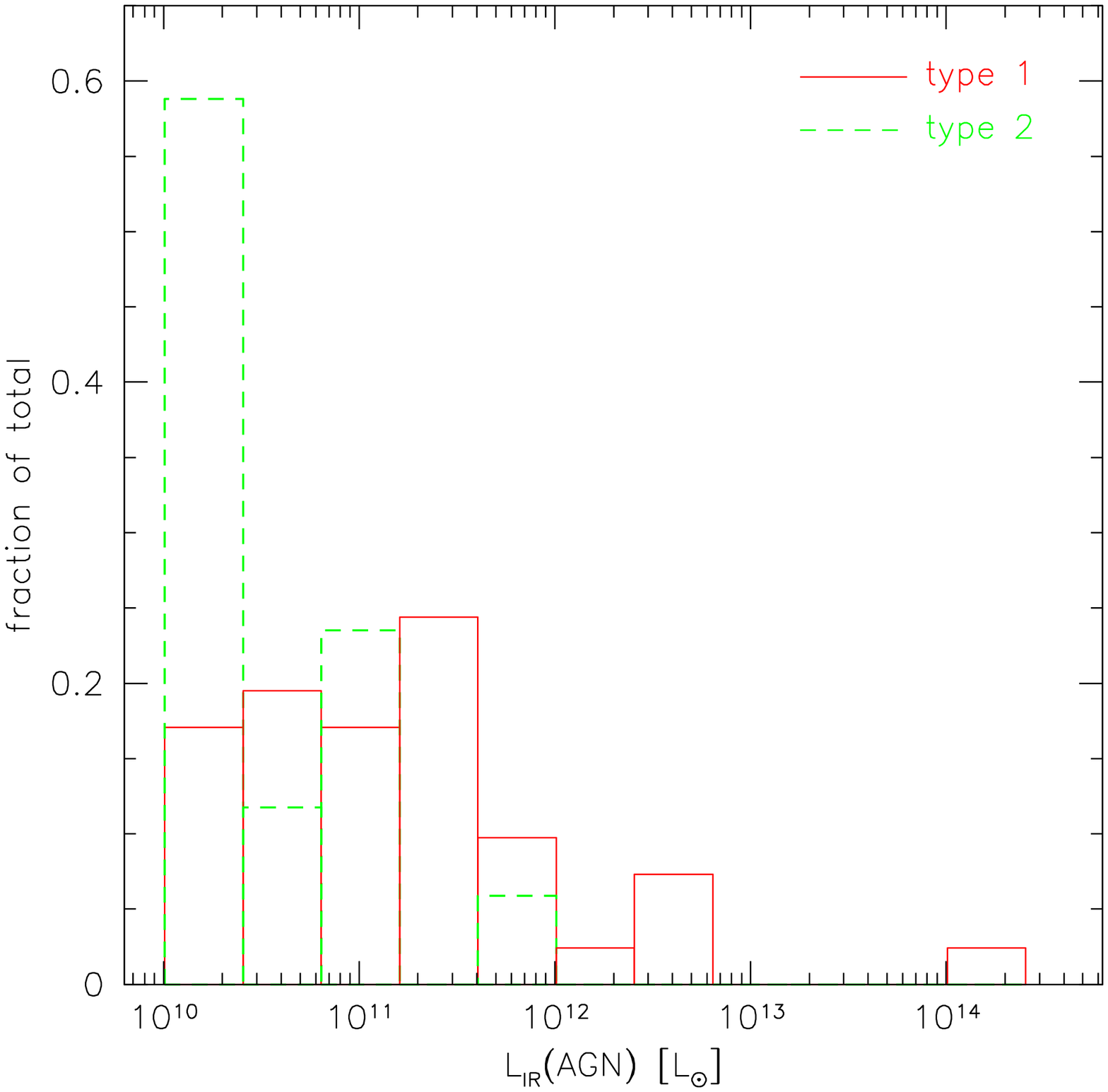} &
\includegraphics[height=.450\textwidth]{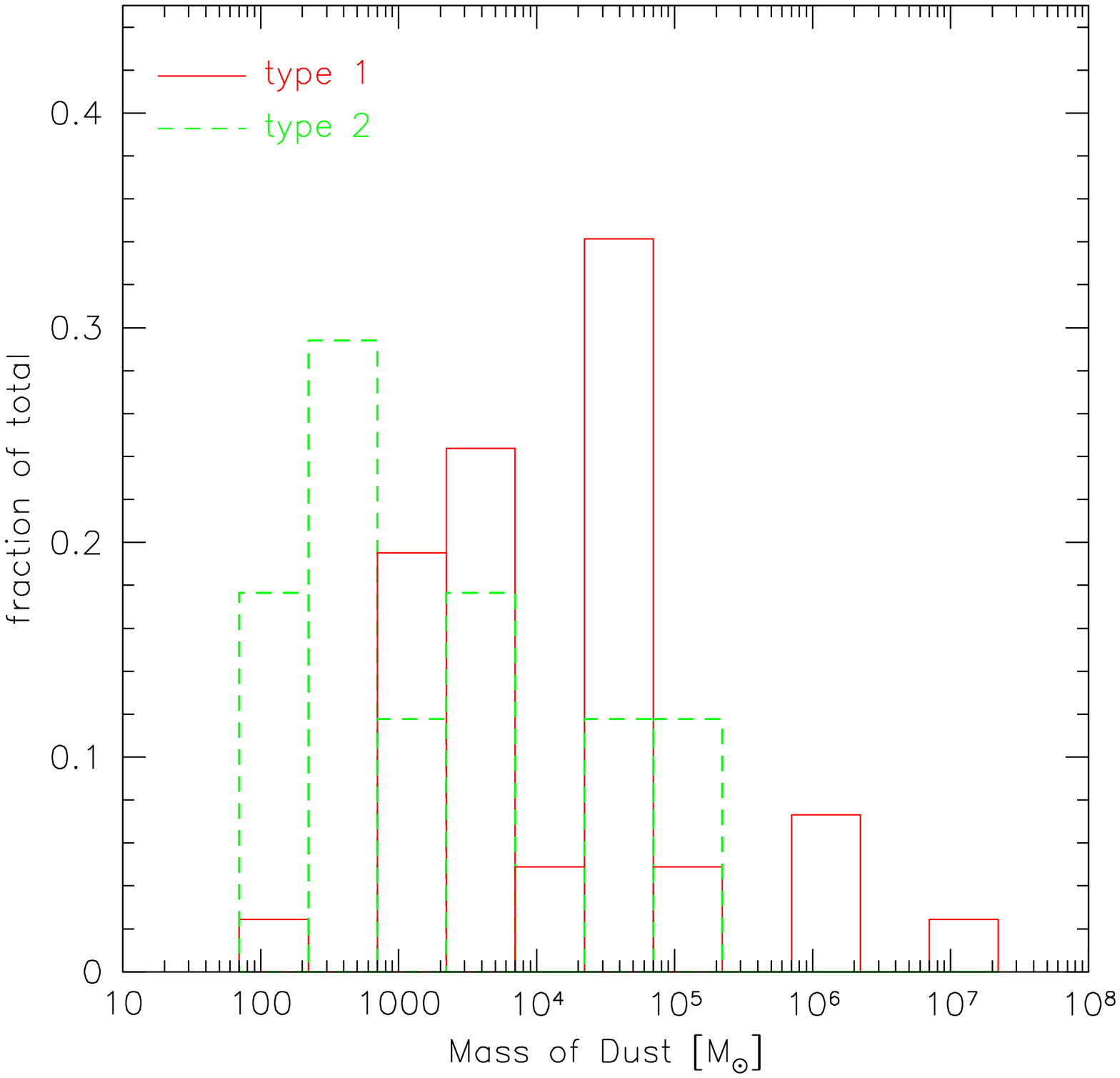} \\
\end{tabular}
\caption{The distribution of the values of dust masses for type-1 and type-2 sources does not seem to highlight deep differences between the two classes, while distribution of the IR luminosity that is ascribed to the AGN component is more peaked towards the low luminosity part of the plot. This effect is due to the fact that the starburst component is more important in type-2 sources.}
\label{fig:props2}
\end{figure*}

As shown in the left panel of Fig. \ref{fig:props}, summarizing a comparison of our
best-fit parameters for the two AGN populations, the average AGN contribution 
to the bolometric IR luminosity is lower than that estimated for the type-1 objects.
We should caution, however, that the two subsamples are distributed quite differently in distance: the redshifts for the type-1 objects (appearing in the same Fig. \ref{fig:props}) are typically higher than for the type-2, due to their easier spectroscopic identification. This bias obviously affects the bolometric AGN luminosity and dust torus mass (Fig. \ref{fig:props2}), while it is unclear at this stage how much it might influence the estimate of the AGN fractional contribution.

It is remarkable that, in spite of this bias, the distributions of the main physical and geometrical parameters for the two classes of AGN are fairly consistent, in particular
the optical depth, with a median value of $\tau(9.7)$ of $6.0$ for the type-2, while it is $10.0$ for the type-1 sample.
In any case, the results should be seen with caution due to the in-homogeneity, incompleteness and random character of our sample.

\section{DISCUSSION AND CONCLUSIONS}
\label{discuss}

This work describes an improved model for the emission of dusty tori around AGN,
including features already occasionally discussed in the literature and
paying particular attention to an accurate computation 
of the radiative transfer solution.
The chosen torus geometry is simple but realistic, a flared torus defined by its 
inner and outer radii and the total opening angle. We adopt the dust grain distribution 
of \cite{MRN}, 
with the density depending on the torus radial coordinate and polar angle. 
The optical depth is computed in great detail, taking into account the different sublimation temperatures for silicate and graphite grains.

The fact that grains with different sizes can have different sublimation 
temperatures may imply a significant complication, especially in the inner 
parts of the torus where the temperature of some dust grains can reach and 
exceed the sublimation limit. While this model accounts for the differences 
between graphite and silicate grains, we assume the sublimation temperature 
to be independent of the grain size. This follows \cite{efstat94}, who have 
shown that this effect is not of much importance especially for high 
densities and optical depths.

Our model is able to reproduce with remarkable accuracy the observed broad-band 
SEDs of a variety of AGN, for which large numbers of 
photometric measurements were available in the literature.
For three well-known nearby active galaxies (Circinus, NGC 1068 and Mrk 231), 
the {\it ISO} mid-IR spectrum was also available, in which cases both the torus 
model and the starburst contributions were strongly constrained. 

\subsection{Spectral fitting of type-1 AGNs}
\label{ty1}

A starburst component is usually required in order to fit the observed SED 
longward of $50$ \mum, as in most cases a pure AGN component is unable to 
reproduce the observed width of the IR bump. From the objects of our sample, only 
four type-1 sources can be fitted by a pure AGN emission 
in the far-IR. The torus emission is in general predominant for such objects, 
while the starburst prevails in type-2 (see Fig. \ref{fig:props}, left panel). 

In this work we pay particular attention to the behaviour of the $9.7$ \mums 
silicate feature, which in many previous publications was found to challenge 
the Unified Model (especially for type-1 objects). The problem was the 
prominence of the feature in absorption for type-2 AGN and its absence in 
type-1 objects, as viewed pole-on. 
This feature arises from dust situated in the innermost regions of the torus,
as seen in Fig. \ref{fig:elements}, were the temperature is close to the sublimation 
limit. Moving far away from the center the feature becomes 
gradually fainter untill it eventually turns into absorption if the optical 
depth is high enough.
Our current results show that a combination of high optical 
depth, moderate torus radius, an either constant density profile
or one radially increasing toward the center, and a proper 
consideration of the lower sublimation temperature of silicate grains, 
reduces the silicate emission or even turns it into slight absorption. 
Under these assumptions, a standard composition of the mixture of dust turns 
out to be entirely compatible with low or no silicate emission in our model.
The addition of a significant starburst contribution with the silicate feature in absorption
tends to further decrease the prominence of the $9.7$ \mums feature in type-1 AGN.

\begin{figure}
\rotatebox{270}{
\includegraphics[height=.5\textwidth]{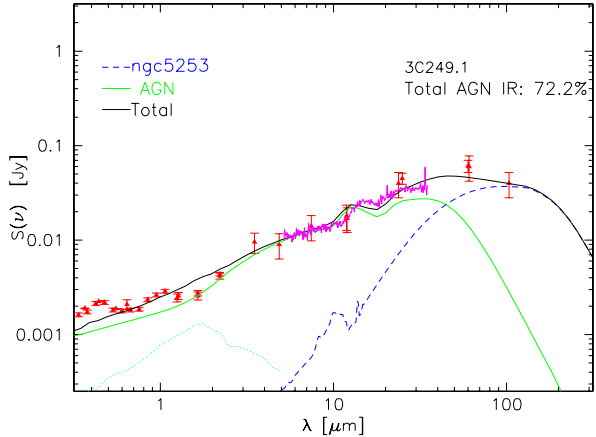}}
\caption{Model fit to the observed SED and the IR Spitzer spectrum (observed by Siebenmorgen et al. 2005) between $5-30$\mums of the luminous QSO 3C249.1. The $9.7$ \mums silicate feature is clearly seen in emission with an intensity which very close to that of our model. The discrepancy in the wavelength of the observed peak is likely due to a different chemical composition of the silicate grain (\citealt{siebenmorgen}; \citealt{leihao}). A non-constant density profile with $\gamma=6$ and $\beta=-1.0$ and a torus aperture angle of $100^\circ$ are used, while the external radius is $\sim 35$ pc.}
\label{fig:3C249.1}
\end{figure}

\begin{figure}
\rotatebox{270}{
\includegraphics[height=.5\textwidth]{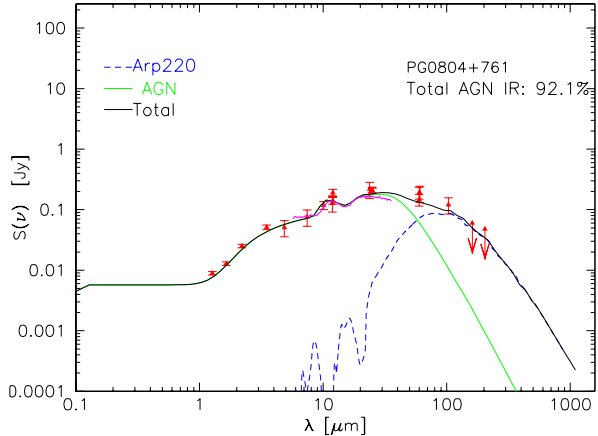}}
\caption{Comparison of the Spitzer spectrum \citep{leihao} and other SED datapoints with our model expectation for the type-1 AGN PG0804+0761. The model is similar to that fitting 3C249.1 (see Fig. \ref{fig:3C249.1} and parameters in Table 2). In this case our fit to the silicate peak is very good.}
\label{fig:PG0804}
\end{figure}

Recent observations with the Spitzer IR space telescope have revealed, 
for the first time, evidence for the silicate feature in moderate emission 
in a few type-1 AGN (\citealt{siebenmorgen}; \citealt{leihao}), with values 
of the $S_{9.7}$ parameter close to unity. Indeed, the latter can be achieved 
with several combinations of our torus parameters (see Fig. \ref{fig:silfeat1}). 
We present in Figs. \ref{fig:3C249.1} and \ref{fig:PG0804} a detailed comparison 
of IR Spitzer spectra, together with SED data in the optical and far-IR, of two 
quasars-1 (3C249.1 and PG0804+761) with evidence for the $9.7$ \mums feature in emission.
We see that our model provides an excellent fit of the spectral shape and 
broad-band data for the latter source. As for the former, the spectrum is 
also well reproduced, except for a slight shift in wavelength probably 
due to a locally different grain mixture \citep{siebenmorgen}. 
In conclusion, we find no need for particular geometries, like a tapered disc, 
or for modified distributions of the silicate grains (clumpy tori) in order 
to suppress the silicate feature in type-1 sources. 

\begin{figure}
\rotatebox{270}{
\includegraphics[height=.5\textwidth]{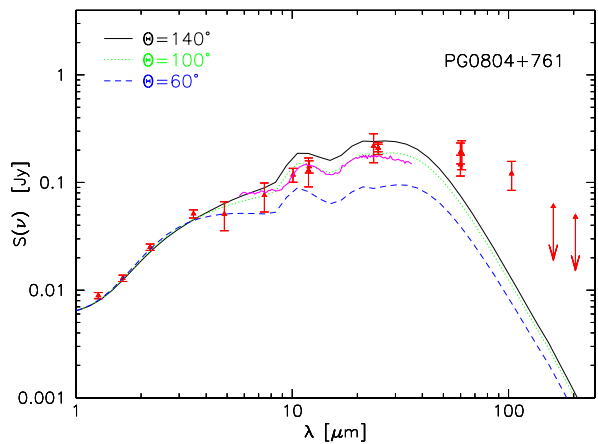}}
\caption{Comparison of predicted spectra for a torus aperture angle of $140^\circ$ (continuous line), $100^\circ$ (dotted line) and $60^\circ$ (dashed line), for a model with the same characteristics as the one used in Fig. \ref{fig:PG0804}. Datapoints as discussed in the caption to Fig. \ref{fig:PG0804}.}
\label{fig:PG0804b}
\end{figure}

\subsection{Torus geometry and the Unified Model}
\label{US}

We now discuss results related to the Unified Scheme. Fig. \ref{fig:props} (right panel) 
illustrates the best-fit values of the parameters reproducing the observed SED of type-1 
and type-2 AGN. Non-constant dust density profiles, in particular a density decreasing 
with the distance from the center, are clearly favored. Furthermore, high values for the 
equatorial optical depths are required in almost all the sources, particularly to 
suppress the $9.7$ \mums emission. Interestingly, in agreement with the predictions
of the Unified Scheme and despite the different average 
distances for the type-1 and type-2 samples, the distributions of the geometrical and 
physical torus parameters do not show significant variation between the two populations. 

Also the distribution of values for the mass of dust in the torus and the luminosity 
of the primary power source presented in Fig. \ref{fig:props2} do not show dramatic 
differences between type-1 and type-2. The slight tendency of type-1 objects to host 
more luminous AGN and more massive tori may be explained with the Malmquist bias 
introduced by the different average distances. More problematic is, instead, the
interpretation of the higher fraction of the AGN contribution in type-1 sources,
seen in Fig. \ref{fig:props} (left panel). It is unclear whether
this is still due to a distance-induced bias, or if it is reflecting an intrinsic 
differentiation of the two populations.

Obviously, not only the silicate feature, but more in general the observed mid-IR 
spectral shape can effectively constrain the intrinsic properties of the 
circum-nuclear dust distribution. An illustration of this is given in Fig. 
\ref{fig:PG0804b}, showing how the spectrum changes by reducing the torus full-opening 
angle $\Theta$ from $140^\circ$ to $100^\circ$ and $60^\circ$: the reduced amount of 
cold dust close to $R_{max}$ for the 60$^\circ$ case makes the spectrum very flat in 
flux density units from $\sim$ 3 to $\sim 50$ \mum, whereas a much steeper one is 
produced by our more standard value of $\Theta=140^\circ$.

A further important implication of our work is that, even accounting for important 
starburst contributions to the longer-wavelength far-IR part, our code was 
essentially unable to obtain good spectral fits to both AGN classes by assuming 
relatively flat tori with low values of the opening angle $\Theta=60^\circ$. 
Instead roughly half of the type-1 AGN and $70$ per cent of the type-2 require 
almost cylindrical tori with $\Theta=140^\circ$. For these, the corresponding
covering factor is $\sim90$ per cent, 
while the remaining sources are better fit with $\Theta=100^\circ$ models, 
corresponding to $\sim75$ percent covering factor. 

Altogether the average covering 
factor for our local sample is close to $80-85$ percent:  4 to 5 obscured AGNs 
are then to be expected on average for each type-1 unobscured object
(we predict very thick circum-nuclear material [$A_V>100$]). 
Since local and high-z AGN samples -- including those selected in the IR 
(e.g. \citealt{rush93}; \citealt{lf04}) or in radio (\citealt{lawrence91}, 
but see also \citealt{maiolino95}) -- do not typically reveal such a high 
incidence of obscured objects, it remains that a large number of completely 
obscured AGN are to be expected, undistinguishable from normal galaxies if 
observed in the optical/near-IR or soft X-rays.
This appears consistent with the results of \cite{franceschini05} from a 
deep combined Spitzer/Chandra survey. As discussed there (see also 
\citealt{maiolino03}; \citealt{gandhi04}; \citealt{fabian03}), an important 
fraction of the obscured quasars and AGN may escape identification, 
even with a good coverage of the mid- and far-IR spectrum, 
as given by the Spitzer data, or deep hard X-ray data.

\subsection{Model degeneracies}
\label{US}

Inevitably, our solutions for the circum-nuclear dust structures in AGNs
suffer some degree of degeneracy in the parameter space.
In our case, the largest source of uncertainty is the
limited spectral coverage, especially in the mid-IR. As already
shown for the three proto-type objects analysed in Sects.
4.1.1, 4.1.2 and 4.1.3, the mid-IR spectrum turns out to be
foundamental constraining the model parameters of AGN dust tori.
The shape of the mid-IR ($1-12$ \mum) continuum is ruled by the optical depth 
and the characteristics of the density law, while the shape of the far-IR 
SED ($20-400$ \mum) is mainly determined by the torus size and 
the amount of cold dust. 

In order to study the degree of degeneracy of our fits,
a detailed case-study of the following three objects was performed:
PG1411+442, as representative of type-1 objects with no evident starburst
contribution, Mrk 1513 as a type-1 source
with significant contribution from starburst emission in the IR, 
and NGC 1614 as a representative of the type-2 sources.

We have compared the observed SEDs for the three sources with all 
model spectra corresponding to the whole parameter set in
Table \ref{tab:mod_grid}, and then considered the solutions 
with a reduced $\chi^2$ below $\chi^2_{best}+3$. 
The accretion luminosity turned out to be the best constrained
quantity, especially in the type-1 sources where the primary 
continuum emission is observable. The scatter was
found to be somewhat larger for the type-2 NGC 1614, in which case it
varied between $\sim 0.75 \times 10^{46}$ and $\sim 10^{46}$
erg/s, the higher values corresponding to model solutions with smaller
torus opening angles. 

The values of the outer radius were fairly constant for the
best fit models in NGC 1614, but differed by up to a factor of
3 in the case of PG1411+442 (both $R_{max}/R_{min}=30$ and $100$
gave acceptable fits). 

Similar results were obtained for the mass of dust. The most
extreme case was again PG1411+442, with a variation by up to a factor
of 4, as a consequence of the $R_{max}/R_{min}$ variation. 
On the other hand, the parameters of the dust density law (equatorial optical
depth and the coefficients $\beta$ and $\gamma$ of the density law)
were found to only slightly vary among the best fits.
The starburst template that better fitted the objects was Arp 220 and was kept
fixed as no acceptable fits could be obtained with other templates.

In conclusion, the available data are not so detailed to allow resolution
of all degeneracies in the torus parameters, which we may understand if we
consider the complexity of the objects under scrutiny and the fact that 
the torus and starburst emissions overlap smoothly in the far-IR.
However, our analysis has shown that significant constraints can be 
achieved about the characteristics of the circum-nuclear dust distributions
in AGNs.

\vspace{0.75cm} \par\noindent   
{\bf ACKNOWLEDGEMENTS} \par  

We want to thank Ralph Siebenmorgen for providing us with the Spitzer Spectrum of 3C249.1.

This research makes use of the NASA/IPAC Extragalactic Database (NED) which
is operated by the Jet Propulsion Laboratory, California Institute of Technology,
under contract with the National Aeronautics and Space Administration.

This work was supported in part by the Spanish Ministerio de
Ciencia y Tecnologia (Grants Nr. PB1998-0409-C02-01 and ESP2002-03716)
and by the EC network "POE'' (Grant Nr. HPRN-CT-2000-00138).

\end{document}